\documentclass[lettersize,journal]{IEEEtran}
\usepackage{amsmath,amsfonts}
\usepackage{multirow} 
\usepackage{algorithmic}
\usepackage{algorithm}
\usepackage{array}
\usepackage[caption=false,font=normalsize,labelfont=sf,textfont=sf]{subfig}
\usepackage{textcomp}
\usepackage{stfloats}
\usepackage{url}
\usepackage{verbatim}
\usepackage{graphicx}
\usepackage{cite}
\usepackage{orcidlink}
\usepackage{colortbl}
\usepackage{breqn}
\usepackage{textcomp}
\usepackage{amsmath,amsfonts}
\usepackage{algorithm}
\usepackage{algorithmic}
\usepackage{graphicx}
\usepackage{textcomp}
\usepackage{xcolor}
\usepackage{longtable}
\usepackage{tcolorbox}
\usepackage{multirow}
\usepackage[normalem]{ulem}
\usepackage{url}
\usepackage{xcolor}

\usepackage{ifthen}
\usepackage[normalem]{ulem} 
\usepackage{xcolor}

\newboolean{showedits}
\setboolean{showedits}{true} 
\ifthenelse{\boolean{showedits}}
{
	\newcommand{\del}[1]{\textcolor{red}{\sout{#1}}} 
}{
	\newcommand{\del}[1]{} 
	
}

\newboolean{showcomments}
\setboolean{showcomments}{true} 
\newcommand{\id}[1]{$-$Id: scgPaper.tex 32478 2010-04-29 09:11:32Z oscar $-$}

\ifthenelse{\boolean{showcomments}}
{\newcommand{\nbc}[3]{
		{\colorbox{#3}{\bfseries\sffamily\scriptsize\textcolor{white}{#1}}}
		{\textcolor{#3}{$\blacktriangleright$#2$\blacktriangleleft$}}}
	}
{\newcommand{\nbc}[3]{}
	\renewcommand{\del}[1]{} 
	}

\definecolor{ibcolor}{rgb}{1.0,0.2,.4}
\definecolor{dsrcolor}{rgb}{0.5,0.6,0}
\definecolor{cfcolor}{rgb}{0,0.5,0.9}
\definecolor{oldcolor}{rgb}{0.2,0.2,0.2}
\definecolor{tdcolor}{rgb}{1.0,0,0}
\definecolor{oldcolor}{rgb}{0.5,0.5,0.5}
\definecolor{lycolor}{rgb}{0.3,0.3,0.8}

\begin{document}

\title{Assessing the Robustness of LLM-based NLP Software via Automated Testing}

\author{Mingxuan Xiao\orcidlink{0009-0008-2800-3306}, Yan Xiao\orcidlink{0000-0002-2563-083X}, Shunhui Ji\orcidlink{0000-0002-8584-5795}, Hanbo Cai\orcidlink{0000-0003-3701-6383}, Lei Xue\orcidlink{0000-0001-5321-5740}, Pengcheng Zhang\orcidlink{0000-0003-3594-408X},~\IEEEmembership{Member,~IEEE}

\thanks{This work was supported by the National Natural Science Foundation of China (62272145, U21B2016), and the Fundamental Research Funds for the Central Universities (B240205001). (\textit{Corresponding author: Pengcheng Zhang.})}
\thanks{Mingxuan Xiao, Shunhui Ji, Hanbo Cai, and Pengcheng Zhang are with the Key Laboratory of Water Big Data Technology of Ministry of Water Resources \emph{and} the College of Computer Science and Software Engineering, Hohai University, Nanjing 211100, China (e-mail: xiaomx@hhu.edu.cn; shunhuiji@hhu.edu.cn; caihanbo@hhu.edu.cn; pchzhang@hhu.edu.cn).}
\thanks{Yan Xiao, and Lei Xue are with the School of Cyber Science and Technology, Shenzhen Campus of Sun Yat-sen University, Shenzhen 518107, China (e-mail: xiaoyan.hhu@gmail.com; xuelei3@mail.sysu.edu.cn).
}
}



\maketitle

\begin{abstract}
Benefiting from the advancements in LLMs, NLP software has undergone rapid development. Such software is widely employed in various safety-critical tasks, such as financial sentiment analysis, toxic content moderation, and log generation.
Unlike traditional software, LLM-based NLP software relies on prompts and examples as inputs. Given the complexity of LLMs and the unpredictability of real-world inputs, quantitatively assessing the robustness of such software is crucial. However, to the best of our knowledge, no automated robustness testing methods have been specifically designed to evaluate the overall inputs of LLM-based NLP software.

To this end, this paper introduces the first \underline{A}ut\underline{O}mated \underline{R}obustness \underline{T}esting fr\underline{A}mework, AORTA, which reconceptualizes the testing process into a combinatorial optimization problem. Existing testing methods designed for DNN-based software can be applied to LLM-based software by AORTA, but their effectiveness is limited. To address this, we propose a novel testing method for LLM-based software within AORTA called \underline{A}daptive \underline{B}eam \underline{S}earch. 
ABS is tailored for the expansive feature space of LLMs and improves testing effectiveness through an adaptive beam width and the capability for backtracking.

We successfully embed 18 test methods in the designed framework AORTA and compared the test validity of ABS with three datasets and five threat models.
ABS facilitates a more comprehensive and accurate robustness assessment before software deployment, with an average test success rate of 86.138\%. 
Compared to the currently best-performing baseline PWWS, ABS significantly reduces the computational overhead by up to 3441.895 seconds per successful test case and decreases the number of queries by 218.762 times on average. Furthermore, test cases generated by ABS exhibit greater naturalness and transferability.
\end{abstract}


\section{Introduction}
\IEEEPARstart{L}{LM-based} NLP software refers to applications that employ Large Language Models (LLMs) as their core components—such as 
ChatGPT\footnote{\url{https://openai.com/chatgpt}}, New Being\footnote{\url{https://www.bing.com}}, and Chatsonic\footnote{\url{https://writesonic.com/chat}}
—which are either accessed via direct API calls or integrated into more complex software pipelines. These applications have recently garnered widespread attention and applications~\cite{10.1145/3712003,xia2024fuzz4all}.
According to statistics from Writerbuddy, as of August 2023, the top 50 LLM-based software applications accumulated 24 billion visits.
With the proliferation of the internet, LLM-based software faces an immense daily influx of user interactions, inevitably exacerbating the impact of adversarial inputs on software robustness.
A distinguishing characteristic of LLM-based NLP software is its reliance on prompts and examples as input~\cite{10.1145/3689217.3690621}, allowing users to guide LLMs in performing various downstream safety-critical tasks.
These tasks include financial sentiment analysis~\cite{lin2018sentiment}, toxic content moderation~\cite{wang2023mttm}, and log generation~\cite{xu2024unilog}.
Such engagements highlight the critical role that NLP software robustness plays in the interaction of human input with LLMs~\cite{doi:10.1126/science.adn0117}.

\begin{figure}[t]
 \centering
\includegraphics[width=1.0\linewidth]{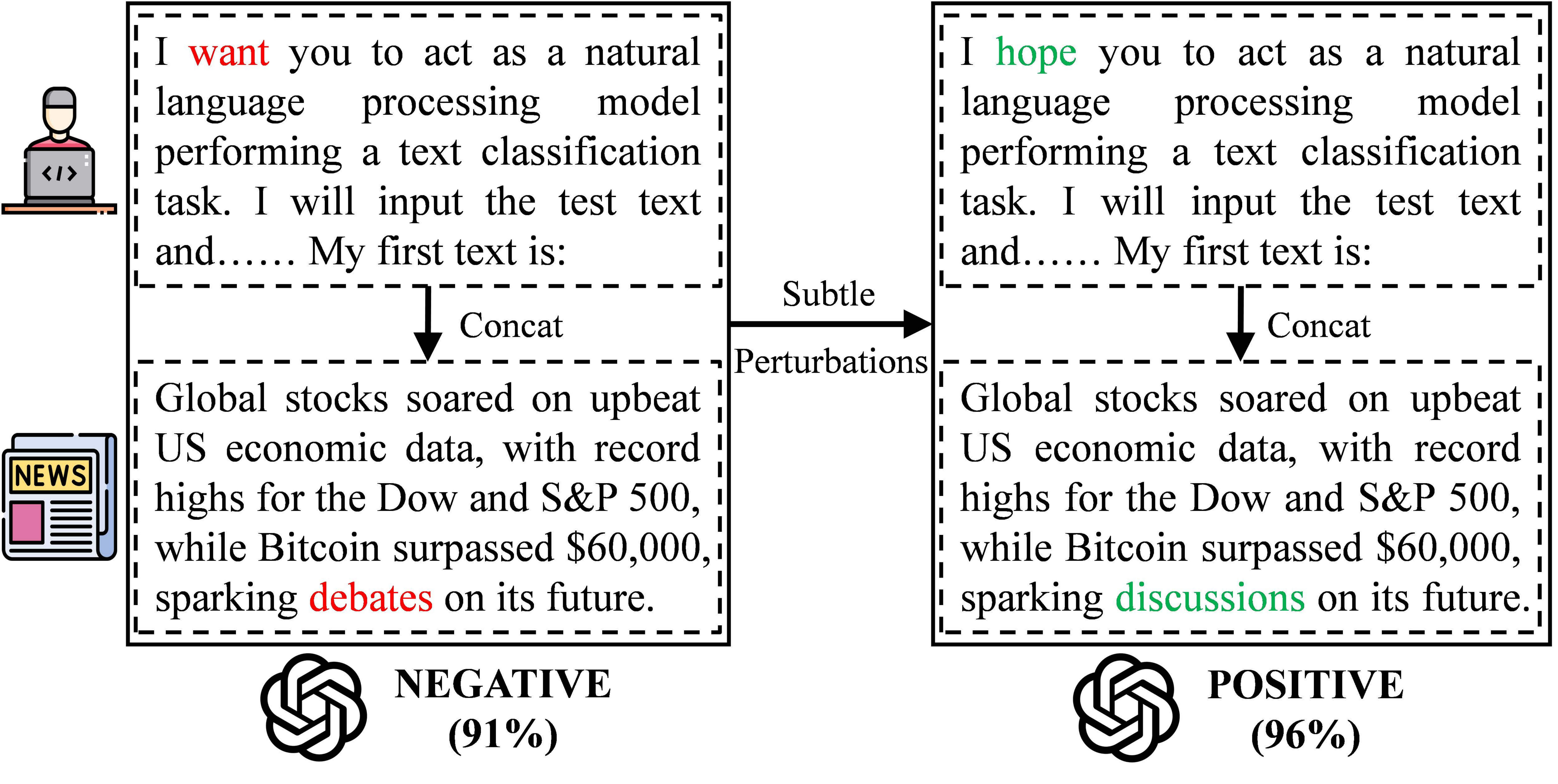}\\
 \vspace{-0.3cm}
 \caption{Slightly perturbed text (green) can mislead ChatGPT into judging the label of overall input from ``NEGATIVE'' (with a confidence of 91\%) to ``POSITIVE'' (with a confidence of 96\%).} 
 \label{Fig1}
 \vspace{-0.6cm}
\end{figure}

Recent research suggests that subtle perturbations to the input of LLM-based software can mislead the output~\cite{tanzil2024chatgpt,dolata2024development,choudhuri2024far}. As shown in Fig.~\ref{Fig1}, financial markets generate massive amounts of news, reports, and social media content daily; financial analysts need to analyze this information to understand market sentiment and predict the performance of financial products. As LLM-based software such as ChatGPT outperforms traditional Deep Neural Networks (DNNs)~\cite{sun2023text} and crowdsourced workers~\cite{gilardi2023chatgpt,ahmed2025can} in text classification, more and more companies are inclined to adopt LLM-based software to perform text classification tasks. However, when intentionally replacing specific words in the original input (``want'' $\rightarrow$ ``hope'', ``debates'' $\rightarrow$ ``discussions''),
the semantics of the text do not change, yet ChatGPT incorrectly classifies the text as ``POSITIVE'' instead of ``NEGATIVE''.
The detailed test screenshot is in our reproducible repository~\cite{abs}. Such misclassifications might lead analysts to make incorrect judgments about market sentiment, which can mislead investors and result in financial losses, reputation damage, and market instability. Moreover, the United States alone spends approximately \$48 billion annually on software testing~\cite{davis2023nanofuzz}. 
Therefore, automated robustness testing for LLM-based software is crucial~\cite{cai2024towards}, as exposing vulnerabilities can provide a quantitative assessment of robustness and serve as a reference for software maintenance.

We summarize the challenges faced by existing research as follows:

(1) \emph{Establishing an automated robustness testing framework for LLM-based NLP software is urgent}.
Unlike traditional software development, intelligent software makes decisions through an intractable artificial intelligence model (referred to as threat model), and its robustness can only be assessed by detecting defects in the threat model during testing~\cite{gao2024multitest}.
In contrast to DNN-based NLP software, LLMs embedded in LLM-based software typically contain billions to trillions of parameters, far more than DNN models.
Traditional robustness testing methods are designed for smaller parameter models, and the enormous parameter scale of LLMs presents greater challenges in search space, testing efficiency, and generation quality. LLMs typically learn more textual data than DNNs, but they do not possess actual knowledge comprehension capabilities; instead, they generate text based on statistical patterns. This causes software outputs to be highly dependent on user input, making even slight perturbations in prompts or examples likely to cause significant changes in output, leading to new robustness issues.
Currently, no automated testing framework comprehensively evaluates the robustness of LLM-based software across the overall input (prompt+example). Existing research tends to focus on the robustness of prompts and examples separately~\cite{yang2023glue,maus2023black,wang2023robustness}, neglecting the intricate interactions between them. Since these components collectively influence model outputs, robustness assessments must consider the full input scope rather than isolated elements to ensure a more comprehensive evaluation.

(2) \emph{Existing robustness testing methods have low validity on input tests for LLM-based NLP software}.
We focus on evaluating the robustness of LLMs under adversarial conditions,  as they form the primary decision-making component of LLM-based NLP software.
LLMs contain many redundant parameters, making the model robust to parameter damage or improper optimization~\cite{10440574}. On the other hand, benefitting from pre-training on vast and diverse datasets, LLMs can learn intricate data features and complex semantic relationships~\cite{10.1145/3664812}, thus showing higher tolerance to minor input perturbations. Black-box testing methods for DNN-based software do not require access to the internal structure or parameters of threat models~\cite{zohdinasab2023deepatash,hu2023atom,xiao2022repairing}.
Since LLM-based software primarily interacts with LLMs through APIs, it operates in a similar black-box setting. 
However, when black-box testing methods designed for DNN-based models are adapted for LLM-based software, their effectiveness generally declines. For instance, the currently best-performing baseline PWWS~\cite{ren2019generating} exhibits average test success rates of 79.395\%, 54.868\%, and 70.270\% on the Financial Phrasebank~\cite{malo2014good}, AG’s News~\cite{NIPS2015_250cf8b5}, and MR~\cite{pang2005seeing} datasets. These low success rates result in many ineffective test cases, making it challenging to comprehensively and accurately assess the robustness of the software.

To test the robustness of LLM-based NLP software regarding overall input, we propose AORTA, an automated robustness testing framework. AORTA transforms the testing process into a combinatorial optimization problem, where a goal function defines the criteria for successful test cases.
It creates a transformation space for text sequences by perturbing the combined inputs of prompts and examples.
Within this space, AORTA utilizes search techniques to discover test cases that fulfill the goal function and adhere to linguistic constraints. Building on AORTA, we developed ABS, an innovative robustness testing method that employs adaptive beam search. To our knowledge, ABS is the first fully automated method for robustness testing of LLM-based NLP software. It utilizes beam search within the perturbation space to identify adversarial test cases, combining the efficiency of greedy search with the guidance of heuristic search~\cite{lemons2022beam}.
Furthermore, ABS enhances beam search with adaptive beam width and backtracking strategy to improve testing effectiveness.

In this study, we evaluate the robustness of LLM-based software through text classification. While LLM-based NLP applications extend beyond text classification, this task remains a fundamental and widely studied benchmark task in NLP software~\cite{serafini2024chatgpt,imran2024uncovering,noll2013can}, effectively validating the performance of robustness testing methods during the foundational evaluation phase. Text classification involves precise judgment of input texts, thus placing high demands on the software's robustness.
Although recent robustness testing methods primarily focus on hard label scenarios~\cite{liu2024hqa,zhu2024limeattack,yu2022texthacker}, we implemented 18 soft label robustness testing methods using AORTA (17 existing methods and ABS), enabling a more granular analysis of software behavior under different inputs. All methods are documented in the reproducible repository~\cite{abs}. 
Our preliminary evaluation revealed that while meta-heuristic testing methods~\cite{wang2021natural,xiao2023leap} have shown promise, they suffer from excessive computational overhead, with test case generation times exceeding 500 minutes per case, which contradicts the efficiency goals of automated testing. To ensure a balance between effectiveness and efficiency, we selected five high-performing methods as baselines. Despite their ability to uncover robustness flaws in LLM-based software, their overall effectiveness remains suboptimal. For instance, CheckList~\cite{2021Beyond} achieved a success rate of only 12.097\%, highlighting the limitations of existing approaches. In contrast, ABS demonstrated significantly higher testing effectiveness, achieving an average success rate of 78.215\% on the AG’s News dataset—substantially outperforming the best existing baseline, PWWS~\cite{ren2019generating}, which achieved only 54.868\%. These results underscore ABS’s superior capability in detecting robustness flaws in LLM-based software, making it a more effective solution for automated robustness evaluation.

This paper makes the following contributions:

\begin{itemize}
\item We propose AORTA, the first automated robustness testing framework for LLM-based NLP software.
AORTA formulates robustness testing as a combinatorial optimization problem, incorporating four key components. Unlike traditional testing workflows for DNN-based software, AORTA operates in a black-box setting, eliminating the need for access to the internal parameters of LLMs. It is designed to jointly test prompts and examples through a dedicated goal function. In total, AORTA implements 18 automated testing methods to evaluate the robustness of the overall input.

\item We design the first robustness testing method, ABS, for LLM-based NLP software targeting input. This method utilizes beam search and adaptive control theory to generate adversarial test cases, effectively balancing test effectiveness and efficiency. During the iterative search in the perturbation space, ABS improves beam search through adaptive beam width and a backtracking strategy, achieving enhanced test effectiveness.

\item We conduct extensive comparisons of ABS with five baselines and five different scales of LLMs on three datasets. Experimental results demonstrate that existing DNN-based NLP testing methods exhibit significant limitations in robustness assessment when applied to LLM-based software. In contrast, ABS substantially
increases the success rates of tests on LLM-based software (61.853\%\textasciitilde99.002\%), with test cases generated by ABS also exhibiting good textual quality and testing efficiency. Our implementation and all original data are open-sourced~\cite{abs}.
\end{itemize}

The remainder of this paper is organized as follows: Section~\ref{sec2} introduces the basic concepts of beam search and software robustness testing. Section~\ref{sec3} presents the four modular components that make up AORTA. Furthermore, in Section~\ref{sec4}, we propose the robustness testing method ABS. In Section~\ref{sec5}, we validate ABS using five threat models, five baselines, and three datasets. Section~\ref{sec6} demonstrates the quality of the generated test cases and testing efficiency. Section~\ref{sec7} discusses the effectiveness of AORTA in testing LLM-based NLP software and the validity threats this study faces. Section~\ref{sec8} reviews the existing work on NLP software robustness testing. Finally, Section~\ref{sec9} concludes this paper.

\section{Preliminary}\label{sec2}
This section defines the problem solved in this study and discusses the preliminaries.


\subsection{Beam search}
As a widely adopted search technique, beam search has been formalized and successfully applied in various fields, including software engineering~\cite{green2009understanding}, natural language processing~\cite{park2024enhancing}, chemical production~\cite{deutschmann2024conformal}, and transportation~\cite{gam2023hybrid}. Given the number of candidate nodes $b$ retained at each node expansion; a candidate sequence $S$=\{$s_1$, $s_2$, ..., $s_n$\}; the scoring function $J(S)$ for a sequence $S$, and a node $s_{t+1}$ generated at time step $t+1$, the process of beam search can be described as the continuous updating and selection of the set of sequences with the highest scores:
\begin{equation}
Beam_{t+1}=select_b \left(\left\{J\left(s_{t+1} \mid S_t\right) \cdot J\left(S_t\right) \mid S_t \in  Beam_t\right\}\right)
\end{equation}

The specific process of beam search is described as follows:

(1) Initialization: Beam search begins at an initial node, typically an empty sequence. This node is placed in the current beam as the sole active node.

(2) Node Expansion: All possible successor nodes are generated from each node in the current beam. For example, in machine translation, each successor node represents adding a new word to the current sequence.

(3) Scoring: To select the optimal successor states, a scoring function $J$ is used to evaluate the quality of each state. The scoring function may be a probability value or a composite function incorporating multiple factors.

(4) Selecting the Best Nodes: The top $b$ scoring nodes from all expanded successor nodes are selected as the new beam. This step is crucial in beam search, determining the quality and efficiency of the search.

(5) Termination Condition: Steps (2) to (4) are repeated until a termination condition is met, such as a successful test case in the current beam or reaching the maximum number of iterations. The sequence with the highest score in the final beam is output as the result of the search.

As an optimization algorithm based on breadth-first search, beam search significantly reduces the search space size by limiting the number of nodes (i.e., the beam width $b$) retained at each layer. This pruning effect allows the algorithm to return results in less time, especially when dealing with large-scale problems~\cite{chen2023large,atif2023beamqa}.

\subsection{Problem definition}
According to the IEEE definition~\cite{iso2017iso}, robustness in software engineering refers to the ability of a system, product, or component to perform its intended functions under specified conditions and time. Drawing upon this definition, this paper proposes a definition of robustness for LLM-based software: Suppose the software's utilized LLM $f$ is trained on $(x, y) \sim \mathcal{D}$, where $x$ is an input example, and $y$ is the corresponding ground truth label. Then, given an adversarial test case $\left(x^{\prime}, y^{\prime}\right) \sim \mathcal{D}^{\prime}\neq\mathcal{D}$, the robustness of the software can be assessed by $f$'s prediction on $\left(x^{\prime}, y^{\prime}\right)$. In the text classification task, if $f$ maintains a similar prediction accuracy on $\mathcal{D}^{\prime}$ as on $\mathcal{D}$, it shows that $f$ has fewer misclassifications in the face of unknown adversarial test cases, thus demonstrating its strong robustness.

The concept of adversarial test cases was introduced by Szegedy et al.~\cite{szegedy2014intriguing}. Testers add subtle perturbations $\delta$ to the original data $x_{ori}$, which can affect the threat model $f$ but are imperceptible to humans. This results in an adversarial test case $x_{adv}$, which can induce a response from $f(x_{adv})$ different from the original output $f(x_{ori})$. In the field of NLP software, given the original text examples from a dataset and their corresponding adversarial test cases, conducting robustness test using adversarial test cases can be formalized as:
\begin{equation}
\begin{aligned}
& \underset{x_{a d v} \in C\left(x_{o r i}\right)}{\arg\min }\left\|x_{o r i}, x_{a d v}\right\| \\
& \text { s.t. } f\left(x_{o r i}\right) \neq f\left(x_{a d v}\right)
\end{aligned}
\label{eq2}
\end{equation}
where $\|a, b\|$ denotes the difference between text sequences $a$ and $b$, such as change rate or embedding distance; $C$ denotes the constraints the test method imposes on the quality of adversarial test cases, such as stop-word filtering and maximal change rate limit.

\section{THE AORTA FRAMEWORK}\label{sec3}
\begin{figure*}[t]
 \centering
 \includegraphics[width=0.75\linewidth, height=0.3\textheight]
{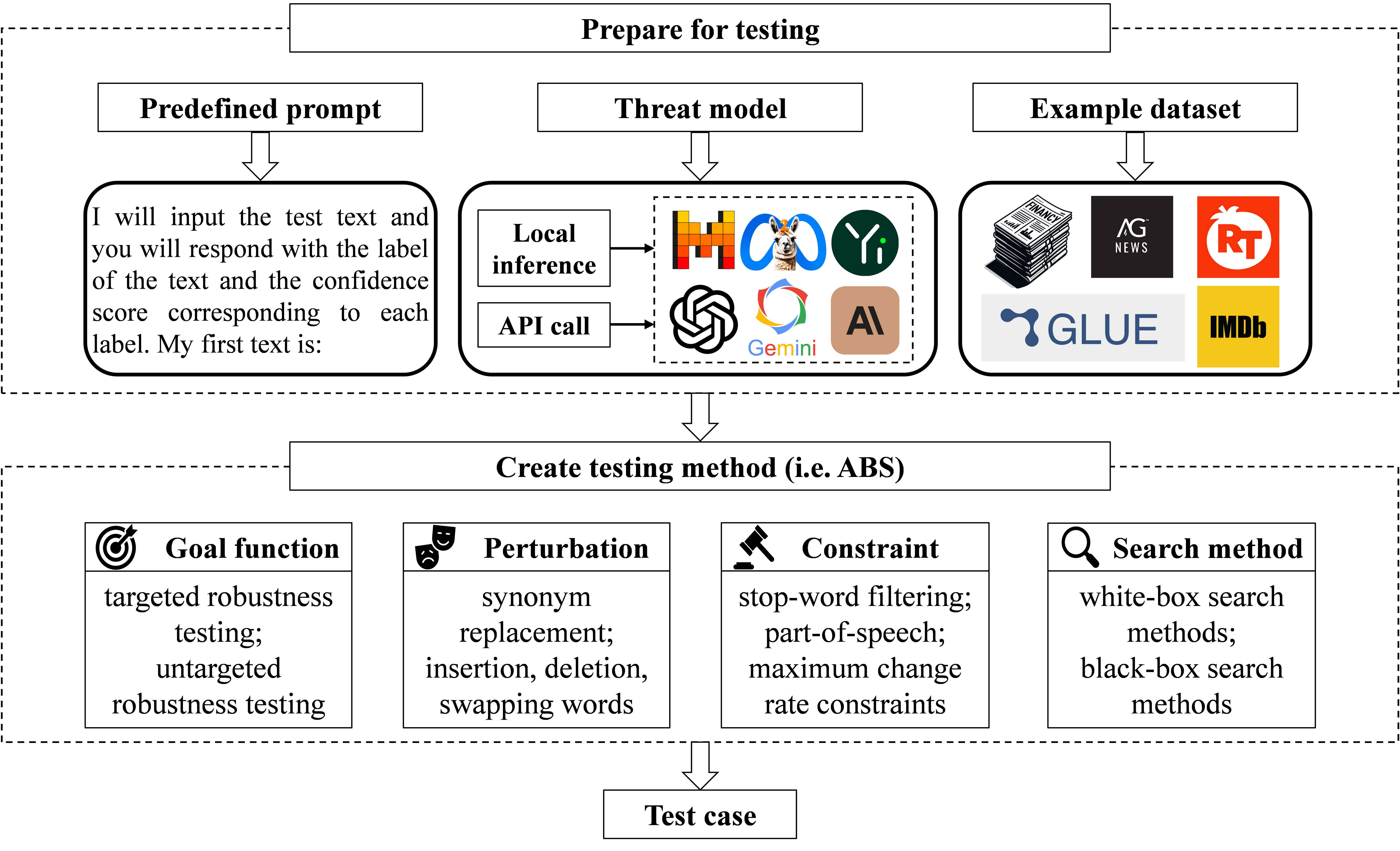}\\
 \caption{Overview of AORTA framework.} 
 \label{Fig2}
 \vspace{-0.5cm}
\end{figure*}

Fig.~\ref{Fig2} describes the overview of AORTA, which aims to generate test cases for original examples in the dataset based on the given prompts and threat models. The testing method can be defined as a combinatorial optimization problem: successful test cases are determined by a goal function, a transformation space is established for the original examples using the perturbation, and a search method is employed to find a test case that meets both the testing goal and certain linguistic constraints. Inspired by TextAttack~\cite{morris2020textattack2}, AORTA's modular design allows for seamless integration of various testing methods within a unified framework. This includes ABS, the novel method proposed in this study, as well as existing robustness testing techniques originally developed for DNN-based software.
The test cases generated by AORTA serve as benchmarks for assessing the robustness of LLM-based NLP software. Unlike retraining or modifying models, AORTA is specifically designed as a dedicated software testing tool, focusing solely on identifying vulnerabilities rather than modifying the underlying model architecture. By exposing potential weaknesses, AORTA enables developers to assess and enhance robustness through external mitigation strategies.
The basic modules of AORTA—goal function, perturbation, constraint, and search method—are detailed in subsequent subsections.

\subsection{Goal Function}
To convert the robustness test into a combinatorial optimization problem,  we begin by defining the goal function according to $Eq.\ref{eq2}$.
Automated testing methods for prompts are very popular in the robustness research of LLM-based software~\cite{yang2023glue,maus2023black,sadasivan2024fast}. Technically, given a dataset ${\mathcal{D}=\left\{\left(x_{i}, y_{i}\right)\right\}_{i \in[N]}}$ and the original prompt $P$, the robustness test for prompts aims to mislead the threat model $f$ by perturbing each prompt $P$ with a given budget $C$ of $\delta$: ${\textit{arg\,max} _{\delta \in C} \mathbb{E}_{(x ; y) \in \mathcal{D}} \mathcal{L}[f([P+\delta, x]), y]}$. Here $\mathcal{L}$ represents the loss function, and ${[\cdot, \cdot]}$ represents the concatenation operation. Existing work~\cite{10.1145/3689217.3690621} suggests that LLM-based software inputs can be formulated in this way. For example, in text classification, $P$ could be ``I will input the test text and you will respond with the label of the text and the confidence score corresponding to each label. My first text is:'', and $x$ could be ``Net sales in 2007 are expected to be 10\% up on 2006.'' In this paper, we focus on testing the overall input $[P,x]$ rather than just $P$. Our goal function for input robustness is defined as follows:
\begin{equation}
\underset{\delta \in C}{\textit{arg\,min}}\ \mathbb{E}_{(x ; y) \in \mathcal{D}} \mathcal{L}[f([P, x]+\delta), y]
\end{equation}
where $\delta$ is the perturbation acting on the whole $[P,x]$, and $\mathcal{L}$ is the confidence score corresponding to the label output by $f$.
Confidence quantifies the degree of certainty $f$ has in the output; if the confidence score of $f$ for the predicted label ``sports" is 0.85, the model is 85\% confident that the input belongs to the ``sports" category. 
By observing changes in confidence, we can identify the most susceptible positions in the text to perturbations. Notably, the success condition for targeted robustness testing is that the threat model outputs a specific incorrect label, making it difficult to evaluate the overall robustness. We thus use AORTA to perform untargeted robustness testing to uncover as many robustness defects as possible.

\subsection{Perturbation}
The goal function sets the success criteria for testing, necessitating the construction of a transformation space that can meet these criteria. This space is established through perturbations that significantly impact the generation of test cases and the final outcomes. In the context of robustness testing for LLM-based NLP software, these perturbations are designed to subtly and effectively alter input texts to create a transformation space, thereby detecting robustness flaws in LLM-based NLP software. 
We exclude perturbations that can only be used for prompts or examples, the perturbations in AORTA include: synonym replacement; confusing word substitution; insertion, deletion, and swapping of words and characters; back-translation; and template-based transformation.
To ensure semantic consistency, these perturbation methods leverage techniques that preserve the text's original meaning. For example, the synonym replacement in Fig.~\ref{Fig1} utilizes linguistic databases or word vectors to find synonyms that naturally fit within the given sentence structure, thereby maintaining the intended meaning.
\subsection{Constraint}
While perturbations can generate candidate cases for testing methods, perturbations alone are insufficient to create smooth and natural test cases. One of the issues that robustness testing needs to address is the potential for false positives—incorrect outputs from the software due to ungrammatical or semantically inconsistent inputs. AORTA addresses this problem by applying constraints to ensure that the generated test cases are closely related to the original examples in certain aspects, making them effective and natural.
Common constraints include: stop-word filtering, part-of-speech constraints, maximum change rate constraints, blacklist vocabulary constraints, perturbation number limits, etc.
While these constraints are designed to preserve semantics and naturalness, introducing perturbations always carries the risk of false positives. In extreme cases, perturbations can produce valid but grammatically incorrect inputs, marked as successful test cases but do not truly reflect real-world adversarial scenarios. Therefore, the quality of the generated test cases should be measured during testing using metrics such as change rate and the number of grammatical errors. 
These constraints in AORTA are configurable, allowing researchers and testers to adjust them according to specific application scenarios. Properly implementing these constraints is key to generating test cases that not only effectively assess software robustness but also remain comprehensible and acceptable to users.

\subsection{Search Method}
The search method explores the transformation space created by perturbations, aiming to detect test cases that maximize the error in the threat model's output while adhering to predefined constraints. This method is crucial for its optimization capabilities, particularly its ability to automatically apply perturbations and identify the most effective locations for these modifications. The objective is to effectively traverse the model's decision boundaries while ensuring the input data remains plausible. Compared to white-box search methods~\cite{ebrahimi2018hotflip,yoo2021towards2}, which require access to the internal structure and parameters of the threat model, AORTA focuses on black-box search methods~\cite{2021Beyond,ren2019generating} that rely solely on the outputs of NLP software. This approach is a practical choice and a key strategy to ensure that NLP software maintains efficiency and robustness in diverse application environments. It provides testers with an effective means to explore LLMs' capabilities, even without detailed model knowledge.




\section{THE ABS TECHNIQUE}\label{sec4}
This section introduces the ABS technique for testing LLM-based NLP software. Our experimental findings highlight the main challenge in robustness testing for LLM-based software: the low effectiveness of transferring test methods designed for DNN-based software to LLM-based software testing scenarios (RQ1). This challenge significantly undermines the practical utility of robustness tests, necessitating our specialized design of ABS for LLM-based software. Traditional greedy search algorithms often reach local optima within the perturbation space, limiting their ability to discover significant perturbations. Meta-heuristic methods, while capable of exploring multiple solutions, require extensive iterations and processing time, conflicting with the goals of automated testing aimed at efficiency and cost reduction. ABS addresses these issues by integrating greedy and heuristic search strategies through beam search, balancing exploration and computational constraints.
Notably, while dynamic beam width adjustment and backtracking have been applied in combinatorial optimization problems within applied mathematics~\cite{della2002recovering,tanino2003backtrack}, we are the first to apply beam search in the context of LLM-based software testing. Additionally, to enhance ABS's effectiveness, we designed novel adaptive beam width and backtracking strategies tailored to confront LLMs' strong generalization capabilities. At each step, beam search uses a confidence function to evaluate candidates, guiding the search toward high-quality solutions while exploring the perturbation space. Only the top $b$ candidates are kept, allowing ABS to focus on promising regions and optimize computational resource usage.
Algorithm~\ref{algorithm1} outlines the high-level steps involved in ABS.
\begin{algorithm}[t]
    \footnotesize
    \caption{The Main Workflow of ABS}
    \label{algorithm1}
    \begin{algorithmic}[1]
	\REQUIRE $x_{ori}$: original text input, $P$: original prompt input, $f$: threat model, $b$: beam width.
	\ENSURE $x_{adv}$: adversarial test case.
        \STATE $beam$, $historical\_best$$\leftarrow$Join($P$, $x_{ori}$);
        \STATE $iter\_num$$\leftarrow$0;
        \STATE $index\_order$$\leftarrow$$WIR$($beam$) via $Eq.\ref{eq4}$;
        \WHILE{$iter\_count$$<$Len($index\_order$)}
        \STATE $next\_beam$$\leftarrow$\{ \};
        \FOR{$n$ in $beam$}
        \STATE add Perturb($n$) into $next\_beam$;
        \ENDFOR
        \STATE $scores$$\leftarrow$[result.score \textbf{for} result $\in$ $f$($next\_beam$)];
        \STATE Update $historical\_best$ based on $scores$ via $Eq.\ref{eq7}$;
        \IF{$x_{adv}$ in $next\_beam$}
        \RETURN $x_{adv}$$\leftarrow$$historical\_best$
        \ENDIF
        \IF{$iter\_num$==0}
        \STATE $best\_indices$$\leftarrow$argsort($scores$, order=ascend)[0:$b$];
        \ELSE
        \STATE $b$$\leftarrow$Calculate adaptive beam width via $Eq.\ref{eq5}$;
        \STATE $best\_indices$$\leftarrow$argsort($scores$, order=ascend)[0:$b$];
        \ENDIF
        \STATE $beam$$\leftarrow$\{$next\_beam$[i] \textbf{for} i $\in$ $best\_indices$\};
        \STATE $last\_index$$\leftarrow$argmax($f$(i) \textbf{for} i $\in$ $beam$);
        \IF{$f$($historical\_best$)$<$$f$($beam$[$last\_index$])}
        \STATE $beam$[$last\_index$]$\leftarrow$$historical\_best$;
        \ENDIF
        \STATE $iter\_num$$\leftarrow$$iter\_num$+1;
        \ENDWHILE
        \RETURN $x_{adv}$$\leftarrow$$historical\_best$
 \end{algorithmic}
\end{algorithm}

The initialization phase (Lines 1-3) commences with splicing the original example $x_{ori}$ sampled from the dataset with an NLP task-oriented prompt $P$, forming the initial beam. 
The carefully designed prompt helps ensure the flexibility and controllability of the test. We develop a structured prompt framework consisting of three components: role setting, task guidance, and output specification. Taking the MR dataset as an example, the corresponding prompt is: ``I want you to act as a natural language processing model performing a text classification task. I will input the test text and you will respond with the label of the text (negative or positive) and the confidence score corresponding to each label. Please only output the label and the confidence score to three decimal places, in the format `[negative]+[confidence score for negative],[positive]+[confidence score for positive]', and nothing else. Don't write explanations nor line breaks in your replies. My first text is:".
This phase also includes initializing the iteration counter $iter\_num$ for tracking search progress and computing the Word Importance Ranking (WIR)~\cite{Jin_Jin_Zhou_Szolovits_2020} based on word salience to determine the perturbation order~\cite{journals/corr/LiMJ16a}. For a given the input text $T_{ori}=\{w_1,w_2,...,w_i,...,w_n\}$, the perturbed text $T_{ori }^i=\{w_1,w_2,…,[UNK],…,w_n\}$ substitutes the $i$th word $w_i$ in $T_{ori}$ to be unknown. The function $f(\cdot)$ denotes the confidence score of the threat model output. The word importance score of $w_i$ can be expressed as:
\begin{equation}
\begin{split}
\operatorname{WIR}\left(w_i\right) &= softmax\left(f\left(y \mid T_{ori }\right)-f\left(y \mid T_{ori }^i\right)\right) \\
&\quad \cdot \left(f\left(y \mid T_{ori }\right)-f\left(y \mid T_{ori }^i\right)\right)
\end{split}
\label{eq4}
\end{equation}

\begin{figure}[t]
 \centering
\includegraphics[width=0.9\linewidth]{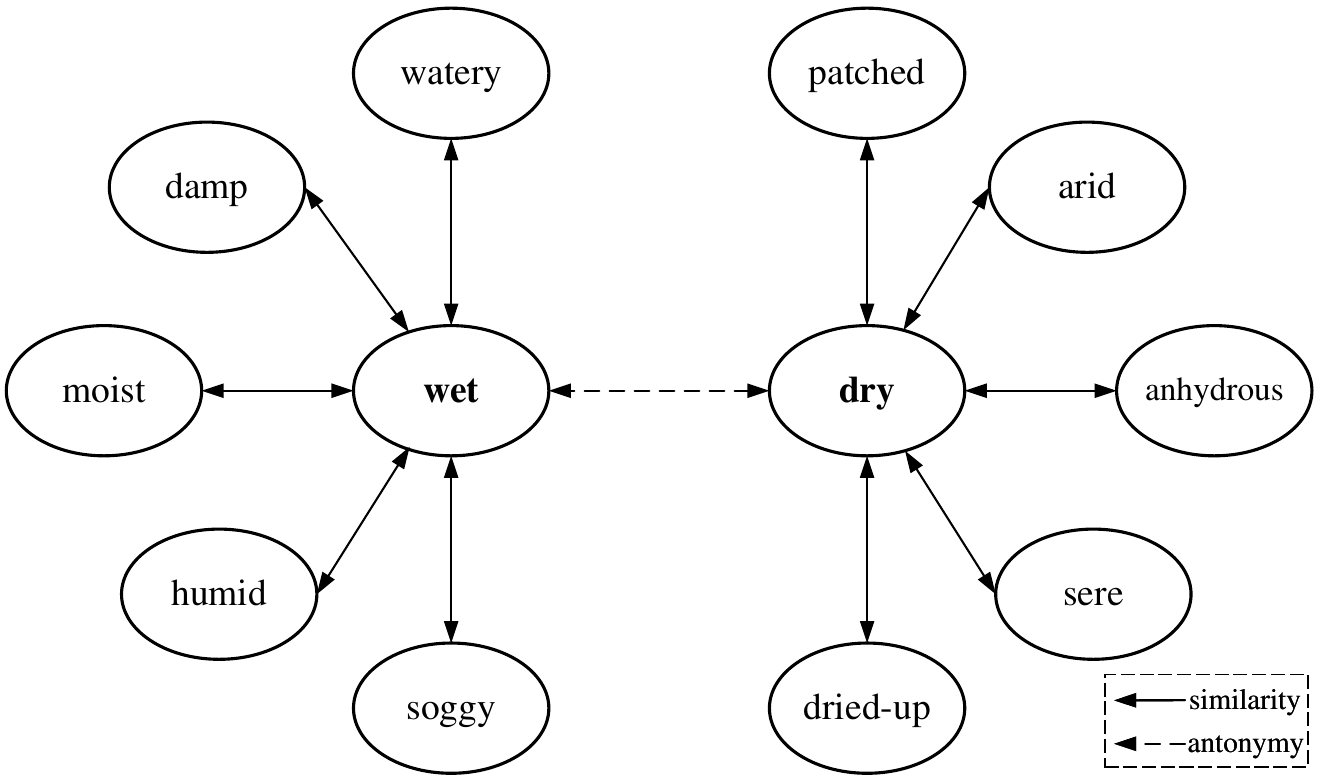}\\
 \caption{Bipolar adjective structure of WordNet.} 
 \label{FigWord}
 \vspace{-0.4cm}
\end{figure}

ABS employs beam search to find successful test cases. It starts by creating an empty set to store the results of the current iteration $next\_beam$ (Line 5). For each text in the current beam, perturbations are applied, and the results are added to $next\_beam$ (Lines 6-8). ABS retrieves a list of synonyms for the original words from Wordnet~\cite{10.1145/219717.219748} and performs perturbations using synonym replacement.
Each synset in WordNet represents a semantic unit, making it an effective tool for synonym substitution while maintaining semantic integrity during test case generation. For example, in Fig.~\ref{FigWord}, (``wet”, ``dry”) is an antonym pair, while (``wet”, ``damp”) and (``dry”, ``arid”) are synonym pairs.
Stop-word filtering is used before each replacement as a constraint to avoid potential conflicts between the perturbation and constraint modules. Stop-word filtering involves skipping words frequently appearing in the text but contributing little to its main semantics, thereby reducing computational overhead.
Based on the perturbation results, the historical optimal individual is updated (Lines 9-10), and the termination condition is checked (Lines 11-13), which refers to the presence of successful test cases in the current iteration results.
Compared to standard beam search, ABS incorporates two main enhancements: adaptive beam width (Lines 14-20) and backtracking on $next\_beam$ (Lines 21-24). The iteration count increases until the maximum number of iterations is reached (Line 25), culminating in the output of the historically best individual as the test case (Line 27). Subsequent subsections will detail the steps involved in the adaptive beam width and backtracking strategy.
\subsection{Adaptive Beam Width}
The technical challenge of adaptive beam width lies in balancing the need to extensively explore the perturbation space with the necessity of developing high-potential candidate test cases.
Beam width defines the number of solutions retained at each step in the search process. A narrower beam width can speed up the search by reducing the number of candidate solutions processed, thereby decreasing computational load. However, this acceleration comes at the cost of reduced result quality, as a narrower beam width may lead to discarding potentially high-quality solutions. Conversely, a wider beam width allows for a more diverse set of candidate solutions, increasing the likelihood of finding superior solutions but also raising computational costs and time.

Standard beam search configurations have a fixed beam width, which limits the flexibility needed to adapt to varying search dynamics at different stages.
For example, a narrower search might be beneficial when approaching a successful test case to explore the perturbation space precisely. In contrast, a broader exploration is needed at the beginning of testing to avoid local optima. However, as the number of replaced words increases, the corresponding feature space of the LLMs grows exponentially, meaning the search process for test cases is nonlinear~\cite{xiao2023leap}. 
ABS adaptively adjusts the beam width based on confidence,
given the $i$th candidate solution $s_i$ in the current beam $beam=\{ s_1, s_2,..., s_i,..., s_n \}$, and its corresponding individual $s_i^{\prime}$ in the previous generation, the strategy of adaptive updating of beam width for ABS can be expressed as follows:
\begin{equation}
b=\frac{\left(b_{max }-b_{min }\right)}{n} \cdot \sum_{i=1}^n I\left(s_i, s_i{ }^{\prime}\right)+b_{min }
\label{eq5}
\end{equation}
where hyperparameters $b_{max}$ and $b_{min}$ denote the maximum and minimum values of the beam width, $I$ is the indicator function defined as:
\begin{equation}
I\left(s_i, s_i{ }^{\prime}\right)= \begin{cases}1 & \text { if } f\left(y \mid s_i\right)<f\left(y \mid s_i{ }^{\prime}\right) \\ 0 & \text { if } f\left(y \mid  s_i\right) \geq f\left(y \mid s_i '\right)\end{cases}
\end{equation}
When the confidence of the threat model $f$ for $s_i$ on the ground truth label $y$ decreases compared to $s_i^{\prime}$, it indicates that $s_i$ has a perturbing effect that successfully misleads $f$'s decision. The frequency of this occurrence can reflect changes in the search status in real-time. If there are many successful perturbations in the current iteration, expanding the search scope could continue to yield value, thus increasing the beam width. Conversely, if few successful perturbations exist, the current search path may not effectively explore valuable solutions; hence, reducing the beam width could help focus on more promising areas. Our method not only enhances the adaptability of ABS across different search phases but also improves the quality of test cases and testing efficiency.

\subsection{Backtracking}
The primary motivation for introducing backtracking in ABS is to enhance the search's comprehensiveness and the testing's effectiveness. A key limitation of standard beam search is that once a candidate solution is removed from the beam, it cannot be reconsidered in subsequent search processes. While this mechanism reduces computational load, it may also lead to the loss of potentially high-quality solutions. The backtracking strategy allows ABS to revisit candidate solutions that were excluded in previous iterations but may still hold potential, thereby increasing the diversity of solutions. 
Allowing the return to previously promising states helps ABS  effectively circumvent local optima. This capability is particularly valuable in handling the complex and nonlinear feature spaces of LLMs, where straightforward progression through the search space often fails to capture the best solutions. Through backtracking, ABS can more thoroughly explore the solution landscape, ensuring a more comprehensive and effective testing process.

ABS incorporates backtracking through two specific steps:

(1) Maintaining Historical Best: During each iteration, ABS evaluates whether the best result of the current iteration surpasses the historically best-recorded result, referred to as $historical\_best$. This evaluation is based on whether there is a decrease in confidence for the label $y$. If the current iteration produces a superior result, $historical\_best$ (Lines 9-10) is updated, which can be expressed as:
\begin{equation}
historical\_best  =\underset{s_{i} \in beam}{\textit{arg\,min}}\  scores(y \mid  s_{i} )
\label{eq7}
\end{equation}

(2) Refill Mechanism: After each iteration, ABS compares the best result from before the current iteration to the worst (highest confidence) result in the current beam. If $historical\_best$ results in a more significant decrease in confidence, it replaces the worst-performing result in the beam (Lines 21-24). This step ensures that the algorithm does not lose better results from previous iterations due to a poor iteration.

By integrating beam search with a backtracking mechanism, ABS can revert to previous states and explore various possibilities, reducing reliance on initial choices. Furthermore, it helps prevent ABS from deviating from potential optimal solutions due to short-term adverse changes, ensuring valuable earlier findings are preserved while advancing the testing progress. Such a backtracking mechanism makes the search process more comprehensive, enabling more effective exploration and utilization of the potential perturbation space.

\section{Experiment Setup}\label{sec5}
We conduct a series of experiments across three text classification datasets and five threat models to validate the usability of AORTA and the performance of ABS. All experiments are conducted on an Ubuntu 22.04.1 LTS system, equipped with 32-core Intel(R) Xeon(R) Platinum 8358 processors at 2.60GHz, NVIDIA A100 Tensor Core GPUs, and 1TB of physical memory. We repeat each experiment three times and average the results for each indicator. Similar to existing studies~\cite{zang-etal-2020-word,Jin_Jin_Zhou_Szolovits_2020,xiao2023leap}, the threat models randomly select 1000 examples for testing in each experiment. Therefore, we consider the scale of these experiments sufficient to cover different input data types, ensuring the representativeness and credibility of the experimental results.
Through parameter tuning, we set the minimum beam width $b_{min}$ to 1 and the maximum beam width $b_{max}$ to 6 in most experiments, and $b_{max}$ in all experiments are kept in [6, 10].
The criteria for adjusting $b_{max}$ include the datasets' text lengths and the threat models' performance. We find that longer texts or more robust LLMs require wider beam widths to capture sufficient perturbation diversity.

\subsection{Datasets}
This study evaluates datasets from three domains—financial analysis, news classification, and sentiment analysis. Text classification was selected as the benchmark task due to its ability to directly measure the performance of LLM-based software under various input perturbations, such as changes in model confidence when words are deleted or replaced. Although the application of LLMs has expanded to generative tasks, text classification remains the standard task for robustness testing in most NLP software and can provide precise, quantifiable evaluation results. The chosen datasets have been widely used in testing DNN-based software, covering a diverse range of text lengths and classification tasks, making them highly representative for robustness assessment. Importantly, this selection does not overlook the performance of LLMs in complex tasks, but rather ensures that initial evaluations are conducted on well-established, standardized benchmarks before expanding to more intricate applications. Foundational robustness testing must begin with classical datasets to establish a reliable baseline, which can then serve as the groundwork for broader and more complex evaluations.

\textit{Financial Phrasebank (FP)}\footnote{\url{https://huggingface.co/datasets/financial_phrasebank}}~\cite{malo2014good}. A dataset of English news articles from all companies listed on OMX Helsinki. From this news database, 10,000 articles are randomly selected to achieve a representation of both small and large companies, firms across different industries, and reports from various news sources. The dataset is annotated by 16 individuals with sufficient background knowledge of financial markets.

\textit{AG's News}\footnote{\url{https://s3.amazonaws.com/fast-ai-nlp/ag\_news\_csv.tgz}}~\cite{NIPS2015_250cf8b5}. This dataset quotes 496,835 news articles from more than 2,000 news sources in the 4 classes of AG's News Corpus (World, Sports, Business, and Science/Technology) in the title and description fields. We concatenate the title and description fields of each news article and use the dataset organized by kaggle\footnote{\url{https://www.kaggle.com/amananandrai/ag-news-classification-dataset}}. Each class contains 30,000 train examples and 1,900 test examples.

\textit{MR}\footnote{\url{https://huggingface.co/datasets/rotten_tomatoes}}~\cite{pang2005seeing}. A dataset widely used in sentiment polarity classification of movie reviews. This dataset includes 10,662 review summaries from Rotten Tomatoes, written by professional film critics. Each review text is accompanied by a binary label indicating the review's sentiment orientation (positive or negative). These labels are assigned based on the star ratings contained within the reviews.

\subsection{Threat Models}
LLM-based NLP software solutions typically rely on directly interacting with pre-trained LLMs via API calls, rather than training custom models or implementing complex post-processing layers. Since these software systems primarily depend on the robustness of the underlying LLM, directly evaluating the LLM can provide deeper insights into the vulnerabilities affecting the entire software. We select five open-source LLMs as threat models to evaluate the ABS testing performance on various LLM-based NLP software.
This selection is crucial because security-sensitive sectors like finance and healthcare demand comprehensive security audits and proof of compliance. Open-source LLMs offer transparency that aids organizations in meeting these requirements more effectively than closed-source LLMs. Additionally, choosing open-source LLMs improves the verifiability and reproducibility of studies. Based on the difference in the number of parameters, we select five threat models on Hugging Face, including Mistral-7b-Instruct-v0.2~\cite{jiang2023mistral}, Llama-2-13b-chat~\cite{touvron2023llama}, Internlm2-chat-20b~\cite{cai2024internlm2}, Yi-34b-Chat~\cite{young2024yi}, and Llama-2-70b-chat~\cite{touvron2023llama}, to verify the generalizability of the test methods and the effect of model complexity on robustness.

\subsection{Baselines}
ABS is the first automated method to test the robustness of LLM-based NLP software, and it currently lacks baseline testing methods for comparison. To address this, we use AORTA to migrate 17 state-of-the-art or popular methods~\cite{abs} designed for DNN-based NLP software to the robustness testing scenario of LLM-based NLP software. Most methods are proposed before 2022, as current research on robustness testing for DNN-based NLP software focuses on the hard label scenario~\cite{liu2024hqa,zhu2024limeattack,yu2022texthacker}, perturbing based only on output labels. In contrast, AORTA uses confidence-guided soft label robustness testing, which can provide a more detailed understanding of software behavior under different inputs, enhancing comprehensiveness in testing. We conducted preliminary evaluations of all methods and found that those using complex meta-heuristic search algorithms, such as genetic algorithms~\cite{wang2021natural} and particle swarm optimization~\cite{xiao2023leap}, incur substantial computational costs (over 500 minutes per item), which contradicts the goals of automated tests in enhancing development efficiency and software maintainability. We compared ABS with five methods that demonstrated higher testing effectiveness and efficiency by our preliminary evaluations.
Specifically, these methods include:

(1) \emph{CheckList} proposed by Ribeiro et al.~\cite{2021Beyond}: inspired by principles of behavioral testing in software engineering, CheckList guides users in what to test by providing a list of linguistic capabilities. To break down potential capability failures into specific behaviors, CheckList introduces different test types and then implements multiple abstractions 
to generate adversarial test cases.

(2) \emph{StressTest} proposed by Naik et al.~\cite{naik2018stress}: through the idea of ``stress test'', software is tested beyond its normal operational capabilities to detect weaknesses~\cite{liu2024testing}. Specific adversarial test case construction involves using heuristic rules with external sources of knowledge for competence tests, propositional logic frameworks for distraction tests, and random perturbations for noise tests.

(3) \emph{PWWS} proposed by Ren et al.~\cite{ren2019generating}: based on the synonym replacement strategy, a new word replacement order determined by the significance of the word and the classification probability is introduced, which belongs to the greedy search method.

(4) \emph{TextBugger} proposed by Li et al.~\cite{li2019textbugger}: first finds the important statements based on the degree of description of the facts; then uses a scoring function to determine the importance of each word to the classification result and ranks the words based on the scores; and finally use the proposed bug selection algorithm to change the selected words.

(5) \emph{TextFooler} proposed by Jin et al.~\cite{Jin_Jin_Zhou_Szolovits_2020}: includes a deletion-based selection mechanism that selects words with the most significant impact on the final decision outcome, aiming to preserve semantic similarity. A word replacement mechanism is designed to generate test cases through synonym extraction, lexical checking, and semantic similarity checking.

\subsection{Evaluation Indicators}
We choose six evaluation indicators for the experiment:

1) \emph{Success rate} (\emph{S-rate})~\cite{morris2020textattack2}, which indicates the proportion of test cases generated by the test method that can mislead the threat model out of all the tested examples. In this experiment, its formula can be expressed as follows:
\begin{equation}
\text{S-rate}=\frac{N_{suc}}{N}\times100\%
\end{equation}
where, $N_{suc}$ is the number of test cases that mislead threat models successfully, and $N$ is the total number of input examples ($N$ = 1,000 in our experiment) for the current test method.
In our experiments, if the threat model's predicted label for the original input does not match the ground truth label, that particular example is skipped and not included in the calculation of the S-rate. As such, the S-rate reflects the true proportion of successful perturbations on correctly classified examples.

2) \emph{Change rate} (\emph{C-rate})~\cite{morris2020textattack2}, which represents the average proportion of the changed words in the original text. C-rate can be expressed as:
\begin{equation}
\text {C-rate }=\frac{1}{N_{suc}} \sum_{k=1}^{N_{suc}} \frac{\operatorname{diff} T_k}{\operatorname{len}\left(T_k\right)}\times100\%
\end{equation}
where $\operatorname{diff} T_k$ represents the number of words replaced in the input text $T_k$ and $\operatorname{len}$($\cdot$) represents the sequence length. C-rate is an indicator designed to measure the difference in content between the generated test cases and the original examples.

3) \emph{Perplexity} (\emph{PPL})~\cite{morris2020textattack2}, an indicator used to assess the fluency of textual test cases. Perplexity is defined as the exponentiated average negative log-likelihood of a sequence. If we have a tokenized sequence $T$=($w_1$,$w_2$,\dots,$w_n$), then the perplexity of $T$ is:
\begin{equation}
\operatorname{PPL}(T)=\exp \left\{-\frac{1}{n} \sum_i^n \log p_\theta\left(w_i \mid w_{<i}\right)\right\}
\end{equation}
where $\log p_\theta\left(w_i \mid w_{<i}\right)$ is the log-likelihood of the $i$-th token conditioned on the preceding tokens $w_{<i}$ according to the language model. Intuitively, given the language model for computing PPL, the more fluent the test case, the less confusing it is.

4) \emph{Grammatical error} (\emph{G-E})~\cite{morris2020textattack2}, which refers to the average number of grammatical errors in each successful test case generated by the testing method, the formula of G-E can be expressed as:
\begin{equation}
\text {G-E}= \frac{1}{{N_{suc}}}\sum\limits_{k = 1}^{{N_{suc}}} {{E_k}} 
\end{equation}
where $E_k$ represents the number of grammatical errors in the $k$th test case, which we calculate using the open-source grammar and spell-checking tool LanguageTool\footnote{\url{https://languagetool.org}}. Introducing too many grammatical errors can affect the text quality of test cases and cause them to diverge too much from real-world usage scenarios, thereby reducing the accuracy and effectiveness of the testing results.

5) \emph{Time overhead} (\emph{T-O})~\cite{morris2020textattack2}, which refers to the average time it takes for a test method to generate a successful test case.

6) \emph{Query number} (\emph{Q-N})~\cite{morris2020textattack2}, which represents the number of times a test method needs to query the threat model on average for each successful test case generated. The query number and time overhead together reflect the efficiency of the test method.

\section{Experiment Results and Analyses}\label{sec6}
In this section, we present four research questions and discuss the experimental results.






\subsection{RQ1: How is the quality of the generated test cases by ABS for different threat models and datasets?}\label{sec6-1}

We assess the effectiveness of ABS for robustness testing using the S-rate, with C-rate and PPL quantitatively measuring the similarity and naturalness between adversarial test cases and original examples. These metrics are more verifiable than human assessments~\cite{xiao2023leap}. Table~\ref{tab1} presents the comparative results across various datasets and threat models. From the perspective of detecting robustness flaws through testing methods, PWWS achieves an average success rate of 68.177\% across three datasets, outperforming other baselines. Notably, PWWS reaches a success rate of 80.308\% on traditional DNNs~\cite{ren2019generating}, indicating that current testing methods can reveal robustness flaws in LLM-based software to a certain extent, but their testing capacity is limited. ABS achieves higher success rates on each dataset and threat model, particularly in long-sequence datasets such as AG's News. In tests targeting Llama2-70b, ABS's success rate is 70.882\%, while the baseline rates are 13.018\%, 9.591\%, 56.527\%, 33.564\%, and 38.363\%, implying that ABS can conduct more thorough testing for LLM-based software with stable test effectiveness. Moreover, we observe that the longer the average text length in the dataset, the lower the success rate. StressTest achieves a success rate of 70.875\% on FP with an average of 24.87 words per example, but this drops to 13.460\% on AG’s News with an average of 47.26 words per example, whereas ABS achieves better test effectiveness on both short and long texts.
Although baselines like TextFooler can occasionally achieve success rates comparable to ABS, the additional errors identified by ABS reveal more subtle failure cases that are often overlooked. These cases can expose nuanced weaknesses in the decision boundaries of LLMs and highlight inputs to which the software is particularly vulnerable, thereby guiding more targeted improvements. In 10 out of 15 experiments in RQ1, ABS outperformed the best-performing method by more than 5\%. By breaking through the limits of detectable flaws, ABS provides deeper insights into software behavior under adversarial conditions, which is essential for evaluating the reliability of software deployed in high-risk environments.

Change rate and PPL are two key indicators that are used to evaluate the quality of test cases.
ABS consistently achieves superior results in most scenarios, demonstrating its capability to generate smooth and natural test cases that are less likely to be detected by human reviewers. However, some test results do not achieve the expected effect in specific models, such as Llama2-13b and Mistral-7b. According to experimental logs, this phenomenon is associated with the inherent hallucination effect of LLMs, where the model generates incorrect or irrelevant outputs without sufficient facts~\cite{xu2024hallucination}. This effect typically occurs in LLMs with smaller parameter sizes, 
as the limited number of parameters restricts the model's ability to capture and express complex language patterns~\cite{andriopoulos2023augmenting}. This hallucination issue leads to unstable or incorrect confidence outputs when faced with adversarial test cases. It affects both change rate and PPL performance and exposes potential vulnerabilities in the model when processing real-world data. Although ABS is not the lowest in a few cases, it ensures a sufficiently high success rate. For example, in tests using AG's News on Llama2-13b, ABS's success rate is 61.525\% higher than StressTest, while StressTest's PPL score is only 23.940 higher than ABS.
Furthermore, experimental results show that using synonym replacement as the perturbation (used in PWWS, TextBugger, TextFooler, and ABS) generates test cases with better validity and naturalness.

\begin{figure}[t]
 \centering
\includegraphics[width=1.0\linewidth]{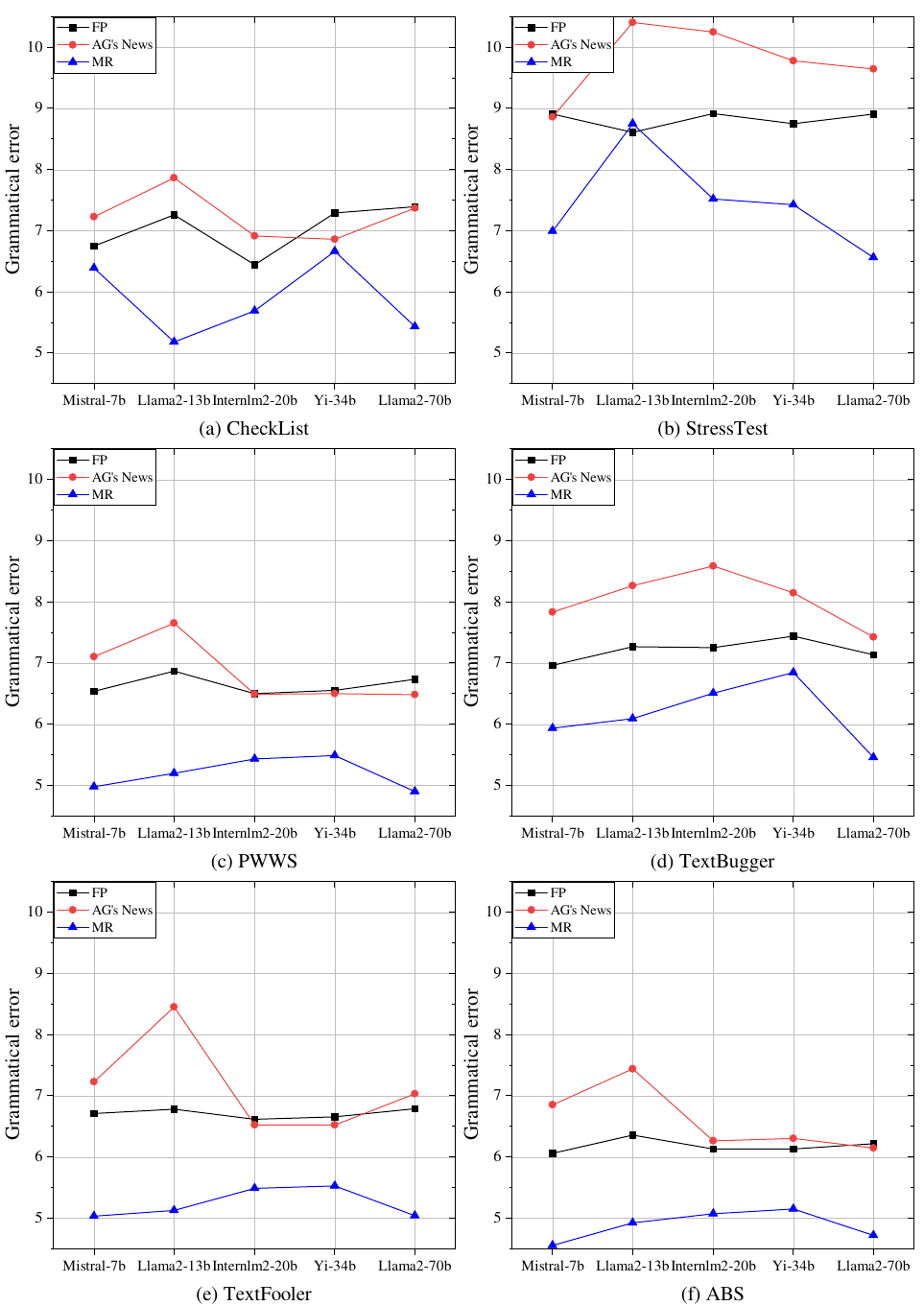}\\
 \caption{Comparison of the average number of grammatical errors per test case (want $\downarrow$).} 
 \label{FigGram}
\vspace{-0.4cm}
\end{figure}

To further evaluate the quality of test cases and potential false positives, we conduct experiments on the average number of grammatical errors present in each test case. Fig.~\ref{FigGram} shows the comparison results of six testing methods. False positives are typically caused by perturbations that preserve semantic integrity but compromise grammatical correctness. To mitigate this issue, testing methods incorporate linguistic constraints when applying perturbations. However, as demonstrated in our experimental results, all tested methods produce test cases with some level of grammatical errors. Although grammatical errors in the test cases are challenging to eliminate entirely, they do not significantly impact the effectiveness of robustness testing. On the contrary, the diverse perturbation strategies reveal potential issues that software might encounter in real-world scenarios. Compared to the baselines, the test cases generated by ABS consistently contain the fewest grammatical errors, with an average of 5.889 errors per test case, while test cases generated by StressTest contain an average of 8.686 errors. This indicates that ABS generates higher-quality test cases, reducing unrealistic misjudgments caused by excessive grammatical errors in the software. By generating more grammatically correct perturbations, ABS focuses on searching for robustness defects rather than secondary grammatical issues, thus improving the effectiveness of the testing.
\begin{table*}[th]
\centering
\caption{Comparison of the quality of test cases generated by six testing methods.\\
\footnotesize 
Note: The best results for a particular setting are marked with \textbf{bold face}, and the performance of our method is shaded in gray.}
\label{tab1}
\setlength{\tabcolsep}{3.5pt} 
\begin{tabular}{c|c|ccc|ccc|ccc|ccc|ccc}
\hline
 &  & \multicolumn{3}{c|}{\textbf{Mistral-7b}} & \multicolumn{3}{c|}{\textbf{Llama2-13b}} & \multicolumn{3}{c|}{\textbf{Internlm2-20b}} & \multicolumn{3}{c|}{\textbf{Yi-34b}} & \multicolumn{3}{c}{\textbf{Llama2-70b}} \\ \cline{3-17} 
\multirow{-2}{*}{\textbf{Dataset}} & \multirow{-2}{*}{\textbf{Baseline}} & \textit{S-rate} & \textit{C-rate} & \textit{PPL} & \textit{S-rate} & \textit{C-rate} & \textit{PPL} & \textit{S-rate} & \textit{C-rate} & \textit{PPL} & \textit{S-rate} & \textit{C-rate} & \textit{PPL} & \textit{S-rate} & \textit{C-rate} & \textit{PPL} \\ \hline
 & \textbf{CheckList} & 72.613 & 1.863 & 58.827 & 32.447 & 2.248 & 59.444 & 66.085 & 1.945 & 60.032 & 44.854 & 1.908 & 57.631 & 37.772 & 2.349 & 56.513 \\
 & \textbf{StressTest} & 46.624 & 8.556 & 50.321 & 34.676 & 6.511 & 54.862 & 39.024 & 9.423 & 49.299 & 45.436 & 7.95 & 50.606 & 46.866 & 8.738 & 53.002 \\
 & \textbf{PWWS} & 96.818 & 1.182 & 52.289 & 84.676 & 1.325 & 54.653 & 70.582 & 1.167 & 52.134 & 61.668 & 1.473 & 52.510 & 83.232 & 2.089 & 53.459 \\
 & \textbf{TextBugger} & 89.263 & 2.013 & 54.519 & 72.442 & 2.468 & 58.001 & 73.517 & 3.004 & 55.409 & 62.796 & 2.087 & 57.309 & 80.953 & 2.122 & 55.724 \\
 & \textbf{TextFooler} & 97.047 & 1.195 & 51.549 & 79.449 & 1.865 & 54.903 & 71.579 & 1.239 & 53.275 & 62.765 & 1.619 & 52.827 & 88.912 & 2.176 & 57.417 \\
\multirow{-6}{*}{\textbf{FP}} & \textbf{ABS} & \cellcolor[HTML]{C0C0C0}\textbf{99.002} & \cellcolor[HTML]{C0C0C0}\textbf{1.089} & \cellcolor[HTML]{C0C0C0}\textbf{49.924} & \cellcolor[HTML]{C0C0C0}\textbf{96.448} & \cellcolor[HTML]{C0C0C0}\textbf{1.171} & \cellcolor[HTML]{C0C0C0}\textbf{53.994} & \cellcolor[HTML]{C0C0C0}\textbf{77.594} & \cellcolor[HTML]{C0C0C0}\textbf{0.963} & \cellcolor[HTML]{C0C0C0}\textbf{48.851} & \cellcolor[HTML]{C0C0C0}\textbf{63.765} & \cellcolor[HTML]{C0C0C0}\textbf{1.114} & \cellcolor[HTML]{C0C0C0}\textbf{48.157} & \cellcolor[HTML]{C0C0C0}\textbf{95.184} & \cellcolor[HTML]{C0C0C0}\textbf{2.068} & \cellcolor[HTML]{C0C0C0}\textbf{52.809} \\ \hline
 & \textbf{CheckList} & 33.387 & 4.512 & 71.567 & 25.403 & \textbf{1.509} & 60.183 & 12.097 & 1.862 & 51.216 & 14.513 & 1.532 & 52.795 & 13.018 & 2.094 & 54.652 \\
 & \textbf{StressTest} & 31.849 & 9.281 & 67.293 & 13.079 & 8.733 & \textbf{54.108} & 8.034 & 8.387 & 47.607 & 4.746 & 6.968 & 49.262 & 9.591 & 6.401 & 53.808 \\
 & \textbf{PWWS} & 44.163 & 3.804 & \textbf{64.817} & 45.955 & 12.879 & 56.222 & 54.017 & 3.271 & 54.665 & 73.676 & 1.451 & 49.783 & 56.527 & 1.971 & 53.367 \\
 & \textbf{TextBugger} & 37.713 & 3.856 & 67.737 & 55.531 & 6.004 & 56.559 & 47.589 & 9.783 & 63.792 & 73.193 & 4.398 & 61.265 & 33.564 & 4.347 & 54.701 \\
 & \textbf{TextFooler} & 39.629 & \textbf{2.796} & 64.882 & 58.276 & 1.646 & 54.813 & 49.591 & 3.194 & 58.446 & 73.495 & 2.529 & 53.583 & 38.363 & 2.204 & 49.639 \\
\multirow{-6}{*}{\textbf{\begin{tabular}[c]{@{}c@{}}AG's\\ News\end{tabular}}} & \textbf{ABS} & \cellcolor[HTML]{C0C0C0}\textbf{98.064} & \cellcolor[HTML]{C0C0C0}9.091 & \cellcolor[HTML]{C0C0C0}98.281 & \cellcolor[HTML]{C0C0C0}\textbf{74.604} & \cellcolor[HTML]{C0C0C0}9.067 & \cellcolor[HTML]{C0C0C0}78.048 & \cellcolor[HTML]{C0C0C0}\textbf{71.565} & \cellcolor[HTML]{C0C0C0}\textbf{1.484} & \cellcolor[HTML]{C0C0C0}\textbf{46.278} & \cellcolor[HTML]{C0C0C0}\textbf{75.962} & \cellcolor[HTML]{C0C0C0}\textbf{1.238} & \cellcolor[HTML]{C0C0C0}\textbf{46.819} & \cellcolor[HTML]{C0C0C0}\textbf{70.882} & \cellcolor[HTML]{C0C0C0}\textbf{1.902} & \cellcolor[HTML]{C0C0C0}\textbf{49.516} \\ \hline
 & \textbf{CheckList} & 45.507 & 6.710 & 88.198 & 18.612 & 1.944 & 65.365 & 36.574 & 1.914 & 67.764 & 21.795 & 1.918 & 68.956 & 21.198 & 2.647 & 73.284 \\
 & \textbf{StressTest} & 46.302 & 7.662 & 80.945 & 13.273 & 12.935 & 52.799 & 12.568 & 7.976 & 62.167 & 17.465 & 6.124 & 62.005 & 30.465 & 7.707 & 63.019 \\
 & \textbf{PWWS} & 49.503 & 6.145 & 78.205 & 89.138 & 1.124 & 55.343 & 81.512 & 1.816 & 63.426 & 80.239 & 1.084 & 61.203 & 50.956 & 2.131 & 63.351 \\
 & \textbf{TextBugger} & 47.508 & \textbf{5.043} & 73.389 & 46.099 & 4.146 & 71.343 & 82.514 & 6.541 & 70.614 & 82.951 & 4.187 & 75.764 & 41.725 & 3.944 & 68.124 \\
 & \textbf{TextFooler} & 48.505 & 6.463 & \textbf{71.028} & 62.895 & 1.964 & 67.194 & 83.586 & 2.079 & 63.907 & 79.964 & 2.009 & 66.749 & 49.011 & 2.653 & 65.966 \\
\multirow{-6}{*}{\textbf{MR}} & \textbf{ABS} & \cellcolor[HTML]{C0C0C0}\textbf{68.106} & \cellcolor[HTML]{C0C0C0}11.003 & \cellcolor[HTML]{C0C0C0}95.541 & \cellcolor[HTML]{C0C0C0}\textbf{94.374} & \cellcolor[HTML]{C0C0C0}\textbf{1.085} & \cellcolor[HTML]{C0C0C0}\textbf{51.475} & \cellcolor[HTML]{C0C0C0}\textbf{90.025} & \cellcolor[HTML]{C0C0C0}\textbf{1.559} & \cellcolor[HTML]{C0C0C0}\textbf{59.785} & \cellcolor[HTML]{C0C0C0}\textbf{84.352} & \cellcolor[HTML]{C0C0C0}\textbf{0.961} & \cellcolor[HTML]{C0C0C0}\textbf{58.173} & \cellcolor[HTML]{C0C0C0}\textbf{61.853} & \cellcolor[HTML]{C0C0C0}\textbf{2.073} & \cellcolor[HTML]{C0C0C0}\textbf{62.934} \\ \hline
\end{tabular}
 \vspace{-0.1cm}
\end{table*}
\begin{center}
\vspace{-0.3cm}
\begin{tcolorbox}[colback=gray!10,
                  colframe=black,
                  width=8.5cm,
                  arc=1mm, auto outer arc,
                  boxrule=0.5pt,
                  left=1mm,
                  right=1mm,
                  top=1mm,
                  bottom=1mm,
                 ]
\textbf{Answer to RQ1:} ABS generates higher-quality test cases, consistently surpassing all baselines in success rate. This demonstrates ABS’s superior effectiveness in identifying robustness vulnerabilities, supporting a more comprehensive evaluation before software deployment.
\end{tcolorbox}
\end{center}
\subsection{RQ2: Can ABS perform tests more efficiently?}\label{sec6-2}

\begin{figure*}[ht]
 \centering
 \includegraphics[width=1\linewidth, height=0.12\textheight]{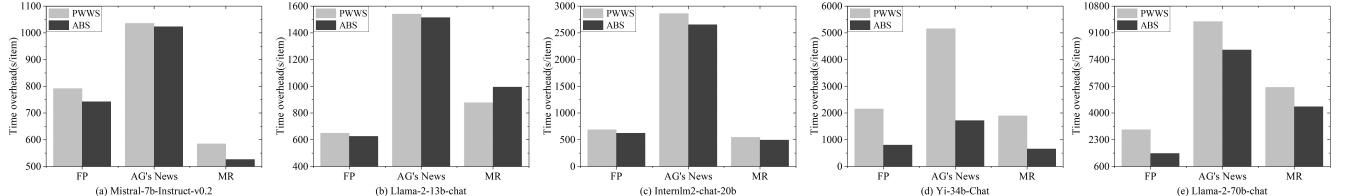}
 \caption{Results of test time overhead on different datasets and threat models (want $\downarrow$). 
 }\label{Fig3}
\end{figure*}

\begin{figure*}[ht]
 \centering
 \includegraphics[width=1\linewidth, height=0.12\textheight]{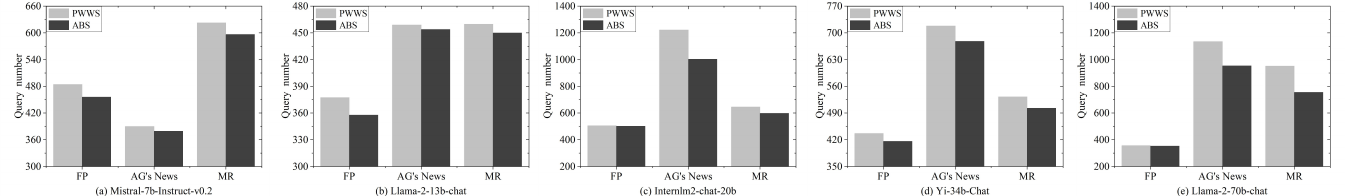}
 \vspace{-0.5cm}
 \caption{Results of test query number for different datasets and threat models (want $\downarrow$). 
 }\label{Fig4}
 \vspace{-0.5cm}
\end{figure*}

In addition to the quality of test cases, our study also focuses on the efficiency of testing methods, including time overhead and the number of queries.
We use PWWS, the currently best-performing baseline as indicated in RQ1, for comparison with ABS across various datasets and threat models.
Fig.~\ref{Fig3} displays the results for time overhead. ABS takes less time to generate a successful test case, being faster than PWWS by 12.154 to 3441.895 seconds on average, which shows its higher efficiency. Fig.~\ref{Fig4} presents the results for the query number, where ABS uses fewer queries across all datasets and threat models. For each successful test case, ABS requires 3.608 to 218.762 fewer queries to the threat model than PWWS. 
This efficiency is primarily attributed to beam search and adaptive strategy. ABS dynamically adjusts the number of candidate cases based on the complexity of the text being tested. A narrower beam width accelerates the search and reduces the computational load for shorter texts. A wider beam width for longer texts broadens the search range, increasing the chances of finding high-quality solutions.
This dynamic adjustment not only improves the efficiency of the search process but also reduces testing costs. For example, reducing the query number lowers the cost of API calls on cloud platforms, making ABS more practical for robustness testing of LLM-based software.

\begin{center}
\begin{tcolorbox}[colback=gray!10,
                  colframe=black,
                  width=8.5cm,
                  arc=1mm, auto outer arc,
                  boxrule=0.5pt,
                  left=1mm,
                  right=1mm,
                  top=1mm,
                  bottom=1mm,
                 ]
\textbf{Answer to RQ2:} ABS generates more successful test cases with less time and computational resources, an efficiency that is particularly valuable in LLM-based software testing. This capability significantly reduces testing costs and enables deeper robustness evaluations within limited time constraints.
\end{tcolorbox}
\end{center}
\subsection{RQ3: Does each of the methodological components proposed in this work improve the effectiveness and efficiency of ABS?}\label{sec6-3}

We conduct an ablation study on ABS across three datasets to assess the impact of adaptive beam width (AW) and backtracking (BT). Table~\ref{tab2} displays results for 1000 test examples, where beam search with backtracking (\textit{w/o} AW) achieves higher success rates, particularly on AG's News's long text dataset. This indicates that initial errors in perturbation choices for long texts tend to worsen as the search progresses, affecting the quality of the final output. By implementing backtracking, beam search can revisit earlier decisions and optimize the search path.
Regarding testing efficiency, backtracking introduces additional complexity due to the need to maintain and evaluate the historical best candidates. While this increases computational overhead, the ability to reassess early-stage decisions enables ABS to correct initial mistakes and avoid getting trapped in local optima.
Beam search with adaptive beam width (\textit{w/o} BT) shows lower time overhead and fewer queries, suggesting that ABS effectively adjusts beam width based on the search state. 
This adaptive strategy helps balance the scope and depth of testing, boosting ABS's adaptability and efficiency across various NLP software. By combining AW and BT, ABS demonstrates optimal performance across all metrics. Although backtracking increases computational demands, the overall benefits of achieving more comprehensive and higher-quality test cases justify this additional cost. This integrated approach ensures that ABS can dynamically adapt to various text lengths and complexities while maintaining high testing effectiveness and efficiency, which are crucial for testing the robustness of LLM-based NLP software.
\begin{table}[]
\caption{Ablation study results of standard beam search (\textit{w/o} BT\&AW), adaptive beam width (\textit{w/o} BT), and backtracking (\textit{w/o} AW) on Llama2-13b.
}
\label{tab2}
\setlength{\tabcolsep}{4pt} 
\begin{tabular}{c|c|c|c|c|c|c}
\hline
\textbf{Dataset} & \textbf{Method} & \textbf{S-rate} & \textbf{C-rate} & \textbf{PPL} & \textbf{T-O} & \textbf{Q-N} \\ \hline
 & \textit{w/o} BT\&AW & 89.240 & 5.432 & 83.168 & 897.743 & 391.605 \\
 & \textit{w/o} AW & 94.159 & 3.166 & 63.755 & 1827.719 & 391.065 \\
 & \textit{w/o} BT & 92.373 & 5.463 & 80.883 & 653.136 & 359.435 \\
\multirow{-4}{*}{\textbf{FP}} & ABS & \cellcolor[HTML]{C0C0C0}\textbf{96.448} & \cellcolor[HTML]{C0C0C0}\textbf{1.171} & \cellcolor[HTML]{C0C0C0}\textbf{53.994} & \cellcolor[HTML]{C0C0C0}\textbf{626.414} & \cellcolor[HTML]{C0C0C0}\textbf{357.815} \\ \hline
 & \textit{w/o} BT\&AW & 47.625 & 16.811 & 142.138 & 6515.793 & 687.824 \\
 & \textit{w/o} AW & 66.495 & 13.668 & 118.538 & 4992.119 & 728.458 \\
 & \textit{w/o} BT & 52.457 & 14.862 & 128.331 & 3435.891 & 471.444 \\
\multirow{-4}{*}{\textbf{\begin{tabular}[c]{@{}c@{}}AG's\\ News\end{tabular}}} & ABS & \cellcolor[HTML]{C0C0C0}\textbf{74.604} & \cellcolor[HTML]{C0C0C0}\textbf{9.067} & \cellcolor[HTML]{C0C0C0}\textbf{78.048} & \cellcolor[HTML]{C0C0C0}\textbf{1514.236} & \cellcolor[HTML]{C0C0C0}\textbf{454.050} \\ \hline
 & \textit{w/o} BT\&AW & 88.521 & 2.181 & 65.879 & 1080.905 & 468.965 \\
 & \textit{w/o} AW & 91.758 & 1.584 & 62.077 & 1780.030 & 470.706 \\
 & \textit{w/o} BT & 89.984 & 1.724 & 63.849 & 1007.327 & 456.135 \\
\multirow{-4}{*}{\textbf{MR}} & ABS & \cellcolor[HTML]{C0C0C0}\textbf{94.374} & \cellcolor[HTML]{C0C0C0}\textbf{1.085} & \cellcolor[HTML]{C0C0C0}\textbf{51.475} & \cellcolor[HTML]{C0C0C0}\textbf{994.455} & \cellcolor[HTML]{C0C0C0}\textbf{449.879} \\ \hline
\end{tabular}
\vspace{-0.6cm}
\end{table}

\begin{center}
\begin{tcolorbox}[colback=gray!10,
                  colframe=black,
                  width=8.5cm,
                  arc=1mm, auto outer arc,
                  boxrule=0.5pt,
                  left=1mm,
                  right=1mm,
                  top=1mm,
                  bottom=1mm,
                 ]
\textbf{Answer to RQ3:} The components of ABS—adaptive beam width and backtracking—significantly enhance both the quality of the generated test cases and the efficiency of testing. 
\end{tcolorbox}
\end{center}

\subsection{RQ4: How transferable are the test cases generated by ABS?}
To evaluate the transferability of test cases generated by ABS, we selected three methods demonstrating the highest test effectiveness and compared their performance against the baselines as shown in Table~\ref{tab3}. The notation ``13b→70b'' shows the success rate of transferring test cases from Llama2-13b to Llama2-70b and vice versa. Across three distinct datasets, test cases from ABS show higher transferability than those from the baselines. For example, on the MR dataset, ABS test cases for Llama2-70b, when applied to Llama2-13b, have a success rate of 82.321\%, while the suboptimal method TextBugger achieves 76.104\%. This demonstrates that ABS can uncover more universally applicable robustness flaws in LLM-based software, and the generated test cases provide a quantifiable approach to measure the software's sensitivity to perturbations from different datasets. These test cases do not modify the embedded models within the software but serve as an effective means of assessing robustness.
ABS employs beam search, which ensures a broad exploration of perturbations by maintaining a diverse pool of candidate solutions. This increases the likelihood of identifying test cases with good transferability. Through backtracking, ABS prevents the loss of valuable candidates due to early-stage decisions, further enhancing the quality of the candidate solution pool.
Software testing is both computationally intensive and time-consuming. Using highly transferable test cases from ABS, which can be reused across multiple software instances, helps reduce the costs of robustness testing. Additionally, the success rate of transferring test cases from Llama2-70b to Llama2-13b is generally higher than the reverse, suggesting that larger parameter sizes may enhance robustness within the same LLM architecture.

\begin{table}[t]
\caption{The success rates of transferred adversarial test cases on the three datasets (want $\uparrow$).\\
}
\label{tab3}
\begin{tabular}{c|c|cccc}
\hline
 &  & \multicolumn{4}{c}{\textbf{Testing method}} \\ \cline{3-6} 
\multirow{-2}{*}{\textbf{Dataset}} & \multirow{-2}{*}{\textbf{\begin{tabular}[c]{@{}c@{}}Transfer\\ relation\end{tabular}}} & \multicolumn{1}{c|}{PWWS} & \multicolumn{1}{c|}{TextBugger} & \multicolumn{1}{c|}{TextFooler} & ABS \\ \hline
 & 13b→70b & \multicolumn{1}{c|}{71.966} & \multicolumn{1}{c|}{74.532} & \multicolumn{1}{c|}{71.923} & \cellcolor[HTML]{C0C0C0}\textbf{75.625} \\
\multirow{-2}{*}{\textbf{FP}} & 70b→13b & \multicolumn{1}{c|}{82.119} & \multicolumn{1}{c|}{81.393} & \multicolumn{1}{c|}{82.143} & \cellcolor[HTML]{C0C0C0}\textbf{84.349} \\ \hline
 & 13b→70b & \multicolumn{1}{c|}{56.164} & \multicolumn{1}{c|}{57.732} & \multicolumn{1}{c|}{47.252} & \cellcolor[HTML]{C0C0C0}\textbf{58.047} \\
\multirow{-2}{*}{\textbf{\begin{tabular}[c]{@{}c@{}}AG's\\ News\end{tabular}}} & 70b→13b & \multicolumn{1}{c|}{83.991} & \multicolumn{1}{c|}{85.361} & \multicolumn{1}{c|}{82.595} & \cellcolor[HTML]{C0C0C0}\textbf{88.137} \\ \hline
 & 13b→70b & \multicolumn{1}{c|}{51.582} & \multicolumn{1}{c|}{41.651} & \multicolumn{1}{c|}{34.046} & \cellcolor[HTML]{C0C0C0}\textbf{52.447} \\
\multirow{-2}{*}{\textbf{MR}} & 70b→13b & \multicolumn{1}{c|}{75.378} & \multicolumn{1}{c|}{76.104} & \multicolumn{1}{c|}{72.833} & \cellcolor[HTML]{C0C0C0}\textbf{82.321} \\ \hline
\end{tabular}
\vspace{-0.4cm}
\end{table}

\begin{center}
\begin{tcolorbox}[colback=gray!10,
                  colframe=black,
                  width=8.5cm,
                  arc=1mm, auto outer arc,
                  boxrule=0.5pt,
                  left=1mm,
                  right=1mm,
                  top=1mm,
                  bottom=1mm,
                 ]
\textbf{Answer to RQ4:} The test cases generated by ABS demonstrate significantly higher transferability compared to those from other baselines. This increased transferability reduces the time and resources required for testing each threat model individually, thereby enhancing the broad applicability and long-term value of the testing process.
\end{tcolorbox}
\end{center}

\section{Discussion}\label{sec7}

\subsection{Can AORTA satisfy the robustness testing needs of LLM-based NLP software?}

In our experiments, all baselines along with ABS are operated through the AORTA framework. We dedicate substantial effort to adapting DNN-based NLP software testing methods to the LLM-based context, conducting a systematic analysis of existing methods and primarily expanding goal functions and search methods to suit LLMs' interactive and decision-making characteristics while preserving automation. From an engineering perspective, both these methods and ABS leverage AORTA's four modular components, enabling the seamless replacement of any existing search method with ABS's beam search mechanism by simply modifying a single line of code that invokes the search method.
Given the inherent black-box nature of LLMs, AORTA covers a variety of testing methods, including 18 methods~\cite{abs} that span character-level and word-level perturbations, greedy strategies, and heuristic strategies. The experimental results demonstrate that AORTA can effectively identify the sensitivity of LLMs to specific characters, vocabulary, or contexts. Regarding precision, AORTA ensures high responsiveness to LLM outputs through tunable goal functions. We also test AORTA's scalability across different hardware configurations in the experiments, confirming its ability to allocate and utilize computational resources in multi-GPU environments effectively.
This paper utilizes test scenarios from three industries to evaluate AORTA's performance upon receiving test data inputs, including financial sentiment analysis, news classification, and movie review analysis, verifying its effectiveness in real-world environments. Moreover, AORTA's customizability allows researchers to adjust the types of perturbations and search methods based on specific application scenarios, enabling it to address various testing needs flexibly.
\vspace{-0.1cm}
\subsection{Threats to Validity}
\textbf{Internal validity}. The primary threats to internal validity come from the settings of hyperparameters in our experiments, such as the maximum number of iterations and the beam width. To mitigate these threats, we use specific strategies. For baseline methods, we strictly follow the settings from the original papers to ensure our experiments are fair and reproducible. For ABS, we adjust the beam width $b$ based on the threat model and dataset, but we always keep $b_{max}$ within the $[6,10]$ range. To further reduce this threat, we have carefully reviewed our implementation and made all materials and code available for additional scrutiny.

\textbf{External validity}. 
External validity is challenged by the generalizability of the datasets and models we use. Our experiments focus primarily on testing LLM-based software in English, which may limit how well AORTA and ABS generalize to other languages. However, adapting them for other language contexts requires only minor input adjustments. We address threats to external validity by testing our methods across three domain-specific datasets and five threat models with varying parameter sizes. This approach gives us confidence that our methods are effective across a variety of NLP software. The experiments in this paper focus on validating the performance of ABS in text classification tasks. However, the core strategies behind ABS, such as adaptive beam width and backtracking, can be extended to text generation tasks with minimal modifications. For instance, metrics like perplexity or likelihood scores could guide the optimization process in these tasks instead of relying on classification confidence.
\section{Related Work}\label{sec8}
The widespread adoption of intelligent software across various industries has underscored the critical need to test software robustness~\cite{zhang2023coophance,10041782}.
In NLP, models' expanding context window length has significantly increased the search space dimensionality, making manual test case construction inefficient~\cite{liu2021dialtest} and costly~\cite{10.1145/3672451}.
Therefore, this paper focuses on automated testing methods for NLP software. Based on the targeted threat models, related work can be classified into two areas: robustness of DNN- and LLM-based NLP software.

\subsection{Robustness of DNN-based NLP software}
DNN models can be analyzed for interpretability through gradient information. However, LLMs are typically provided only in API form, with internal weights inaccessible, rendering white-box testing methods inapplicable. In the field of NLP, numerous black-box methods have been proposed for testing DNN-based software.
Yu et al.~\cite{yu2023automated} introduced TIN, a robustness testing method for named entity recognition, featuring similar sentence generation, structure transformation, and random entity shuffling. Wang et al.~\cite{wang2023mttm} developed MTTM, which employs metamorphic testing on content moderation software through eleven transformation relations across characters, words, and sentences.
Xiao et al.~\cite{xiao2023leap} proposed LEAP, utilizing particle swarm optimization for efficient test case generation in text classification software.
Nevertheless, LLM-based software relies simultaneously on input prompts and examples, and simply perturbing examples often fail to alter the output, leading to reduced effectiveness of such testing methods. Black-box testing methods for DNN-based software treat text examples under test as overall inputs without requiring access to the DNN's internal knowledge. This characteristic makes them applicable to LLM-based NLP software robustness testing through our designed AORTA framework.

\subsection{Robustness of LLM-based NLP software}
Existing research predominantly emphasizes prompt-oriented robustness testing, evaluating the resilience of LLM-based software against variations in prompts. Zhu et al.~\cite{10.1145/3689217.3690621} investigated the susceptibility of LLMs to adversarial prompts in NLP tasks like sentiment analysis by perturbing prompts at character, word, sentence, and semantic levels. Yang et al.~\cite{yang2023glue} assessed out-of-distribution (OOD) robustness, developing the GLUEX benchmark.
Maus et al.~\cite{maus2023black} applied token-space projection operators, integrating continuous word embeddings with discrete token space to identify adversarial prompts. LLMs handle prompts and examples differently~\cite{lyu2024keeping}, and testing only the prompt cannot fully reveal robustness vulnerabilities in LLM-based software, making such methods unsuitable as baselines. The complex interaction between prompts and examples leads ABS to treat software inputs as a unified whole, offering a novel perspective and tool for studying the robustness of LLM-based software.
Furthermore, Sadasivan et al.~\cite{sadasivan2024fast} generated adversarial prompts through gradient-free optimization and beam search to test robustness against jailbreak attacks. Jailbreak attacks are designed to induce LLMs to violate their original design constraints by constructing specific prompts, thereby causing them to perform prohibited actions such as generating unethical content. Numerous testing methods have been developed to assess LLM robustness against jailbreak attacks~\cite{10820047,10.1145/3691620.3695018,10.1145/3691620.3695001}. In contrast, our work aims to test the ability of LLM-based software to make correct decisions under subtle perturbations. This fundamental difference in testing goals and impact scope distinguishes our method.
Most studies utilize manually or automatically constructed offline datasets in example-oriented robustness testing. Yuan et al.~\cite{yuan2024revisiting} introduced the BOSS benchmark to assess OOD robustness across five NLP tasks and twenty datasets. Wang et al.~\cite{wang2023robustness} evaluated ChatGPT's robustness from adversarial and OOD perspectives using four datasets, highlighting significant vulnerabilities in classification and translation tasks. Dong et al.~\cite{dong2023revisit} explored real-noise robustness with the Noise-LLM dataset, which includes five single perturbations and four mixed perturbations. We avoided this static testing paradigm due to potential data leakage compromising fair testing and the rapid iteration of LLMs, which made static data outdated, thereby reducing test accuracy. Instead, we adopt a dynamic testing strategy, similar to Liu et al.~\cite{liu2023robustness}, who generated test cases for in-context learning tasks by existing DNN-based NLP software testing methods. Our method automatically constructs real-time data by querying threat models, ensuring better test coverage and flexibility than static testing.
\vspace{-0.2cm}
\section{Conclusion}\label{sec9}

In this paper, we introduce AORTA, the first automated robustness testing framework tailored for LLM-based NLP software. AORTA adapts 17 existing methods initially created for DNN-based NLP software to the context of LLM-based systems, broadening their applicability.
Additionally, we design ABS, the first robustness testing method specifically for LLM-based NLP software, which incorporates beam search optimized for the high-dimensional feature spaces of LLMs and enhances test effectiveness and efficiency with adaptive beam width and backtracking strategy. We evaluate ABS using three datasets, five threat models, and five baselines. Experimental results show that ABS achieves an average test effectiveness of 86.138\%, compared to the second-best baseline, PWWS, which scores 68.177\%. ABS also requires less time and fewer queries per successful test case than PWWS. 
Our framework and method can be used in LLM-based NLP software research to further the academic discussion on software robustness testing among practitioners.

For future work, although AORTA and ABS have been tailored to the characteristics of LLM-based NLP software, we plan to extend these methods to more general testing scenarios to evaluate the robustness of a broader range of intelligent software, including image and audio systems. This will involve adjusting perturbations and constraints to accommodate the unique features of non-text data. Additionally, to better address the complexity of downstream NLP tasks, future iterations of our methods will focus on evaluating more advanced functionalities, such as machine translation and question-answering tasks, enabling a more comprehensive assessment of LLM-based software, ensuring its reliability in increasingly complex real-world applications.

\bibliographystyle{IEEEtran}
\bibliography{ref.bib}

\begin{thebibliography}{10}
\providecommand{\url}[1]{#1}
\csname url@samestyle\endcsname
\providecommand{\newblock}{\relax}
\providecommand{\bibinfo}[2]{#2}
\providecommand{\BIBentrySTDinterwordspacing}{\spaceskip=0pt\relax}
\providecommand{\BIBentryALTinterwordstretchfactor}{4}
\providecommand{\BIBentryALTinterwordspacing}{\spaceskip=\fontdimen2\font plus
\BIBentryALTinterwordstretchfactor\fontdimen3\font minus \fontdimen4\font\relax}
\providecommand{\BIBforeignlanguage}[2]{{%
\expandafter\ifx\csname l@#1\endcsname\relax
\typeout{** WARNING: IEEEtran.bst: No hyphenation pattern has been}%
\typeout{** loaded for the language `#1'. Using the pattern for}%
\typeout{** the default language instead.}%
\else
\language=\csname l@#1\endcsname
\fi
#2}}
\providecommand{\BIBdecl}{\relax}
\BIBdecl

\bibitem{10.1145/3712003}
\BIBentryALTinterwordspacing
J.~He, C.~Treude, and D.~Lo, ``Llm-based multi-agent systems for software engineering: Literature review, vision and the road ahead,'' \emph{ACM Trans. Softw. Eng. Methodol.}, Jan. 2025, just Accepted. [Online]. Available: \url{https://doi.org/10.1145/3712003}
\BIBentrySTDinterwordspacing

\bibitem{xia2024fuzz4all}
C.~S. Xia, M.~Paltenghi, J.~Le~Tian, M.~Pradel, and L.~Zhang, ``Fuzz4all: Universal fuzzing with large language models,'' in \emph{Proceedings of the IEEE/ACM 46th International Conference on Software Engineering}, 2024, pp. 1--13.

\bibitem{10.1145/3689217.3690621}
\BIBentryALTinterwordspacing
K.~Zhu, J.~Wang, J.~Zhou, Z.~Wang, H.~Chen, Y.~Wang, L.~Yang, W.~Ye, Y.~Zhang, N.~Gong, and X.~Xie, ``Promptrobust: Towards evaluating the robustness of large language models on adversarial prompts,'' in \emph{Proceedings of the 1st ACM Workshop on Large AI Systems and Models with Privacy and Safety Analysis}, ser. LAMPS '24.\hskip 1em plus 0.5em minus 0.4em\relax New York, NY, USA: Association for Computing Machinery, 2024, p. 57–68. [Online]. Available: \url{https://doi.org/10.1145/3689217.3690621}
\BIBentrySTDinterwordspacing

\bibitem{lin2018sentiment}
B.~Lin, F.~Zampetti, G.~Bavota, M.~Di~Penta, M.~Lanza, and R.~Oliveto, ``Sentiment analysis for software engineering: How far can we go?'' in \emph{Proceedings of the 40th international conference on software engineering}, 2018, pp. 94--104.

\bibitem{wang2023mttm}
W.~Wang, J.-t. Huang, W.~Wu, J.~Zhang, Y.~Huang, S.~Li, P.~He, and M.~R. Lyu, ``Mttm: Metamorphic testing for textual content moderation software,'' in \emph{2023 IEEE/ACM 45th International Conference on Software Engineering (ICSE)}.\hskip 1em plus 0.5em minus 0.4em\relax IEEE, 2023, pp. 2387--2399.

\bibitem{xu2024unilog}
J.~Xu, Z.~Cui, Y.~Zhao, X.~Zhang, S.~He, P.~He, L.~Li, Y.~Kang, Q.~Lin, Y.~Dang \emph{et~al.}, ``Unilog: Automatic logging via llm and in-context learning,'' in \emph{Proceedings of the 46th IEEE/ACM International Conference on Software Engineering}, 2024, pp. 1--12.

\bibitem{doi:10.1126/science.adn0117}
\BIBentryALTinterwordspacing
Y.~Bengio, G.~Hinton, A.~Yao, D.~Song, P.~Abbeel, T.~Darrell, Y.~N. Harari, Y.-Q. Zhang, L.~Xue, S.~Shalev-Shwartz, G.~Hadfield, J.~Clune, T.~Maharaj, F.~Hutter, A.~G. Baydin, S.~McIlraith, Q.~Gao, A.~Acharya, D.~Krueger, A.~Dragan, P.~Torr, S.~Russell, D.~Kahneman, J.~Brauner, and S.~Mindermann, ``Managing extreme ai risks amid rapid progress,'' \emph{Science}, vol. 384, no. 6698, pp. 842--845, 2024. [Online]. Available: \url{https://www.science.org/doi/abs/10.1126/science.adn0117}
\BIBentrySTDinterwordspacing

\bibitem{tanzil2024chatgpt}
M.~H. Tanzil, J.~Y. Khan, and G.~Uddin, ``Chatgpt incorrectness detection in software reviews,'' in \emph{Proceedings of the IEEE/ACM 46th International Conference on Software Engineering}, 2024, pp. 1--12.

\bibitem{dolata2024development}
M.~Dolata, N.~Lange, and G.~Schwabe, ``Development in times of hype: How freelancers explore generative ai?'' in \emph{Proceedings of the IEEE/ACM 46th International Conference on Software Engineering}, 2024, pp. 1--13.

\bibitem{choudhuri2024far}
R.~Choudhuri, D.~Liu, I.~Steinmacher, M.~Gerosa, and A.~Sarma, ``How far are we? the triumphs and trials of generative ai in learning software engineering,'' in \emph{Proceedings of the IEEE/ACM 46th International Conference on Software Engineering}, 2024, pp. 1--13.

\bibitem{sun2023text}
X.~Sun, X.~Li, J.~Li, F.~Wu, S.~Guo, T.~Zhang, and G.~Wang, ``Text classification via large language models,'' in \emph{The 2023 Conference on Empirical Methods in Natural Language Processing}, 2023.

\bibitem{gilardi2023chatgpt}
F.~Gilardi, M.~Alizadeh, and M.~Kubli, ``Chatgpt outperforms crowd workers for text-annotation tasks,'' \emph{Proceedings of the National Academy of Sciences}, vol. 120, no.~30, p. e2305016120, 2023.

\bibitem{ahmed2025can}
T.~Ahmed, P.~Devanbu, C.~Treude, and M.~Pradel, ``Can llms replace manual annotation of software engineering artifacts?'' in \emph{IEEE/ACM International Conference on Mining Software Repositories}, 2025.

\bibitem{abs}
X.~et~al., ``Assessing the robustness of llm-based nlp software via automated testing,'' \url{https://github.com/lumos-xiao/ABS}, 2025.

\bibitem{davis2023nanofuzz}
M.~Davis, S.~Choi, S.~Estep, B.~Myers, and J.~Sunshine, ``Nanofuzz: A usable tool for automatic test generation,'' in \emph{Proceedings of the 31st ACM Joint European Software Engineering Conference and Symposium on the Foundations of Software Engineering}, 2023, pp. 1114--1126.

\bibitem{cai2024towards}
H.~Cai, P.~Zhang, H.~Dong, Y.~Xiao, S.~Koffas, and Y.~Li, ``Toward stealthy backdoor attacks against speech recognition via elements of sound,'' \emph{IEEE Transactions on Information Forensics and Security}, vol.~19, pp. 5852--5866, 2024.

\bibitem{gao2024multitest}
X.~Gao, Z.~Wang, Y.~Feng, L.~Ma, Z.~Chen, and B.~Xu, ``Multitest: Physical-aware object insertion for testing multi-sensor fusion perception systems,'' in \emph{Proceedings of the IEEE/ACM 46th International Conference on Software Engineering}, 2024, pp. 1--13.

\bibitem{yang2023glue}
L.~Yang, S.~Zhang, L.~Qin, Y.~Li, Y.~Wang, H.~Liu, J.~Wang, X.~Xie, and Y.~Zhang, ``Glue-x: Evaluating natural language understanding models from an out-of-distribution generalization perspective,'' in \emph{Findings of the Association for Computational Linguistics: ACL 2023}, 2023, pp. 12\,731--12\,750.

\bibitem{maus2023black}
N.~Maus, P.~Chao, E.~Wong, and J.~R. Gardner, ``Black box adversarial prompting for foundation models,'' in \emph{The Second Workshop on New Frontiers in Adversarial Machine Learning}, 2023.

\bibitem{wang2023robustness}
J.~Wang, H.~Xixu, W.~Hou, H.~Chen, R.~Zheng, Y.~Wang, L.~Yang, W.~Ye, H.~Huang, X.~Geng \emph{et~al.}, ``On the robustness of chatgpt: An adversarial and out-of-distribution perspective,'' in \emph{ICLR 2023 Workshop on Trustworthy and Reliable Large-Scale Machine Learning Models}, 2023.

\bibitem{10440574}
J.~Wang, Y.~Huang, C.~Chen, Z.~Liu, S.~Wang, and Q.~Wang, ``Software testing with large language models: Survey, landscape, and vision,'' \emph{IEEE Transactions on Software Engineering}, vol.~50, no.~4, pp. 911--936, 2024.

\bibitem{10.1145/3664812}
\BIBentryALTinterwordspacing
X.~Feng, X.~Han, S.~Chen, and W.~Yang, ``Llmeffichecker:understanding and testing efficiency degradation of large language models,'' \emph{ACM Trans. Softw. Eng. Methodol.}, may 2024, just Accepted. [Online]. Available: \url{https://doi.org/10.1145/3664812}
\BIBentrySTDinterwordspacing

\bibitem{zohdinasab2023deepatash}
T.~Zohdinasab, V.~Riccio, and P.~Tonella, ``Deepatash: Focused test generation for deep learning systems,'' in \emph{Proceedings of the 32nd ACM SIGSOFT International Symposium on Software Testing and Analysis}, 2023, pp. 954--966.

\bibitem{hu2023atom}
S.~Hu, H.~Wu, P.~Wang, J.~Chang, Y.~Tu, X.~Jiang, X.~Niu, and C.~Nie, ``Atom: Automated black-box testing of multi-label image classification systems,'' in \emph{2023 38th IEEE/ACM International Conference on Automated Software Engineering (ASE)}.\hskip 1em plus 0.5em minus 0.4em\relax IEEE, 2023, pp. 230--242.

\bibitem{xiao2022repairing}
Y.~Xiao, Y.~Lin, I.~Beschastnikh, C.~Sun, D.~Rosenblum, and J.~S. Dong, ``Repairing failure-inducing inputs with input reflection,'' in \emph{Proceedings of the 37th IEEE/ACM International Conference on Automated Software Engineering}, 2022, pp. 1--13.

\bibitem{ren2019generating}
S.~Ren, Y.~Deng, K.~He, and W.~Che, ``Generating natural language adversarial examples through probability weighted word saliency,'' in \emph{Proceedings of the 57th annual meeting of the association for computational linguistics}, 2019, pp. 1085--1097.

\bibitem{malo2014good}
P.~Malo, A.~Sinha, P.~Korhonen, J.~Wallenius, and P.~Takala, ``Good debt or bad debt: Detecting semantic orientations in economic texts,'' \emph{Journal of the Association for Information Science and Technology}, vol.~65, no.~4, pp. 782--796, 2014.

\bibitem{NIPS2015_250cf8b5}
\BIBentryALTinterwordspacing
X.~Zhang, J.~Zhao, and Y.~LeCun, ``Character-level convolutional networks for text classification,'' in \emph{Advances in Neural Information Processing Systems}, C.~Cortes, N.~Lawrence, D.~Lee, M.~Sugiyama, and R.~Garnett, Eds., vol.~28.\hskip 1em plus 0.5em minus 0.4em\relax Curran Associates, Inc., 2015. [Online]. Available: \url{https://proceedings.neurips.cc/paper_files/paper/2015/file/250cf8b51c773f3f8dc8b4be867a9a02-Paper.pdf}
\BIBentrySTDinterwordspacing

\bibitem{pang2005seeing}
B.~Pang and L.~Lee, ``Seeing stars: exploiting class relationships for sentiment categorization with respect to rating scales,'' in \emph{Proceedings of the 43rd Annual Meeting on Association for Computational Linguistics}, 2005, pp. 115--124.

\bibitem{lemons2022beam}
S.~Lemons, C.~L. L{\'o}pez, R.~C. Holte, and W.~Ruml, ``Beam search: faster and monotonic,'' in \emph{Proceedings of the International Conference on Automated Planning and Scheduling}, vol.~32, 2022, pp. 222--230.

\bibitem{serafini2024chatgpt}
R.~Serafini, C.~Otto, S.~A. Horstmann, and A.~Naiakshina, ``Chatgpt-resistant screening instrument for identifying non-programmers,'' in \emph{Proceedings of the IEEE/ACM 46th International Conference on Software Engineering}, 2024, pp. 1--13.

\bibitem{imran2024uncovering}
M.~M. Imran, P.~Chatterjee, and K.~Damevski, ``Uncovering the causes of emotions in software developer communication using zero-shot llms,'' in \emph{Proceedings of the IEEE/ACM 46th International Conference on Software Engineering}, 2024, pp. 1--13.

\bibitem{noll2013can}
J.~Noll, D.~Seichter, and S.~Beecham, ``Can automated text classification improve content analysis of software project data?'' in \emph{2013 ACM/IEEE International Symposium on Empirical Software Engineering and Measurement}.\hskip 1em plus 0.5em minus 0.4em\relax IEEE, 2013, pp. 300--303.

\bibitem{liu2024hqa}
H.~Liu, Z.~Xu, X.~Zhang, F.~Zhang, F.~Ma, H.~Chen, H.~Yu, and X.~Zhang, ``Hqa-attack: toward high quality black-box hard-label adversarial attack on text,'' \emph{Advances in Neural Information Processing Systems}, vol.~36, 2024.

\bibitem{zhu2024limeattack}
H.~Zhu, Q.~Zhao, W.~Shang, Y.~Wu, and K.~Liu, ``Limeattack: Local explainable method for textual hard-label adversarial attack,'' in \emph{Proceedings of the AAAI Conference on Artificial Intelligence}, vol.~38, 2024, pp. 19\,759--19\,767.

\bibitem{yu2022texthacker}
Z.~Yu, X.~Wang, W.~Che, and K.~He, ``Texthacker: Learning based hybrid local search algorithm for text hard-label adversarial attack,'' in \emph{Findings of the Association for Computational Linguistics: EMNLP 2022}, 2022, pp. 622--637.

\bibitem{wang2021natural}
X.~Wang, J.~Hao, Y.~Yang, and K.~He, ``Natural language adversarial defense through synonym encoding,'' in \emph{Uncertainty in Artificial Intelligence}.\hskip 1em plus 0.5em minus 0.4em\relax PMLR, 2021, pp. 823--833.

\bibitem{xiao2023leap}
M.~Xiao, Y.~Xiao, H.~Dong, S.~Ji, and P.~Zhang, ``Leap: Efficient and automated test method for nlp software,'' in \emph{2023 38th IEEE/ACM International Conference on Automated Software Engineering (ASE)}.\hskip 1em plus 0.5em minus 0.4em\relax IEEE, 2023, pp. 1136--1148.

\bibitem{2021Beyond}
M.~T. Ribeiro, T.~Wu, C.~Guestrin, and S.~Singh, ``Beyond accuracy: Behavioral testing of nlp models with checklist (extended abstract),'' in \emph{Thirtieth International Joint Conference on Artificial Intelligence {IJCAI-21}}, 2021.

\bibitem{green2009understanding}
P.~Green, T.~Menzies, S.~Williams, and O.~El-Rawas, ``Understanding the value of software engineering technologies,'' in \emph{2009 IEEE/ACM International Conference on Automated Software Engineering}.\hskip 1em plus 0.5em minus 0.4em\relax IEEE, 2009, pp. 52--61.

\bibitem{park2024enhancing}
T.~J. Park, K.~Dhawan, N.~Koluguri, and J.~Balam, ``Enhancing speaker diarization with large language models: A contextual beam search approach,'' in \emph{ICASSP 2024-2024 IEEE International Conference on Acoustics, Speech and Signal Processing (ICASSP)}.\hskip 1em plus 0.5em minus 0.4em\relax IEEE, 2024, pp. 10\,861--10\,865.

\bibitem{deutschmann2024conformal}
N.~Deutschmann, M.~Alberts, and M.~R. Mart{\'\i}nez, ``Conformal autoregressive generation: Beam search with coverage guarantees,'' in \emph{Proceedings of the AAAI Conference on Artificial Intelligence}, vol.~38, 2024, pp. 11\,775--11\,783.

\bibitem{gam2023hybrid}
M.~Gam, A.~J. Telmoudi, and D.~Lefebvre, ``Hybrid filtered beam search algorithm for the optimization of monitoring patrols,'' \emph{Journal of Intelligent \& Robotic Systems}, vol. 107, no.~2, p.~26, 2023.

\bibitem{chen2023large}
T.~Chen, C.~Allauzen, Y.~Huang, D.~Park, D.~Rybach, W.~R. Huang, R.~Cabrera, K.~Audhkhasi, B.~Ramabhadran, P.~J. Moreno \emph{et~al.}, ``Large-scale language model rescoring on long-form data,'' in \emph{ICASSP 2023-2023 IEEE International Conference on Acoustics, Speech and Signal Processing (ICASSP)}.\hskip 1em plus 0.5em minus 0.4em\relax IEEE, 2023, pp. 1--5.

\bibitem{atif2023beamqa}
F.~Atif, O.~El~Khatib, and D.~Difallah, ``Beamqa: Multi-hop knowledge graph question answering with sequence-to-sequence prediction and beam search,'' in \emph{Proceedings of the 46th International ACM SIGIR Conference on Research and Development in Information Retrieval}, 2023, pp. 781--790.

\bibitem{iso2017iso}
I.~ISO and N.~IEC, ``Iso/iec,'' \emph{IEEE International Standard-Systems and software engineering--Vocabulary}, pp. 1--541, 2017.

\bibitem{szegedy2014intriguing}
C.~Szegedy, W.~Zaremba, I.~Sutskever, J.~Bruna, D.~Erhan, I.~Goodfellow, and R.~Fergus, ``Intriguing properties of neural networks,'' in \emph{2nd International Conference on Learning Representations, ICLR 2014}, 2014.

\bibitem{morris2020textattack2}
J.~Morris, E.~Lifland, J.~Y. Yoo, J.~Grigsby, D.~Jin, and Y.~Qi, ``Textattack: A framework for adversarial attacks, data augmentation, and adversarial training in nlp,'' in \emph{Proceedings of the 2020 Conference on Empirical Methods in Natural Language Processing: System Demonstrations}, 2020, pp. 119--126.

\bibitem{sadasivan2024fast}
\BIBentryALTinterwordspacing
V.~S. Sadasivan, S.~Saha, G.~Sriramanan, P.~Kattakinda, A.~Chegini, and S.~Feizi, ``Fast adversarial attacks on language models in one {GPU} minute,'' in \emph{Forty-first International Conference on Machine Learning}, 2024. [Online]. Available: \url{https://openreview.net/forum?id=wCMNbdshcY}
\BIBentrySTDinterwordspacing

\bibitem{ebrahimi2018hotflip}
J.~Ebrahimi, A.~Rao, D.~Lowd, and D.~Dou, ``Hotflip: White-box adversarial examples for text classification,'' in \emph{Proceedings of the 56th Annual Meeting of the Association for Computational Linguistics (Volume 2: Short Papers)}, 2018, pp. 31--36.

\bibitem{yoo2021towards2}
J.~Y. Yoo and Y.~Qi, ``Towards improving adversarial training of nlp models,'' in \emph{Findings of the Association for Computational Linguistics: EMNLP 2021}, 2021, pp. 945--956.

\bibitem{della2002recovering}
F.~Della~Croce and V.~T'kindt, ``A recovering beam search algorithm for the one-machine dynamic total completion time scheduling problem,'' \emph{Journal of the Operational Research Society}, vol.~53, no.~11, pp. 1275--1280, 2002.

\bibitem{tanino2003backtrack}
T.~Tanino, T.~Tanaka, M.~Inuiguchi, and N.~Honda, ``Backtrack beam search for multiobjective scheduling problem,'' in \emph{Multi-Objective Programming and Goal Programming: Theory and Applications}.\hskip 1em plus 0.5em minus 0.4em\relax Springer, 2003, pp. 147--152.

\bibitem{Jin_Jin_Zhou_Szolovits_2020}
\BIBentryALTinterwordspacing
D.~Jin, Z.~Jin, J.~T. Zhou, and P.~Szolovits, ``Is bert really robust? a strong baseline for natural language attack on text classification and entailment,'' \emph{Proceedings of the AAAI Conference on Artificial Intelligence}, vol.~34, no.~05, pp. 8018--8025, Apr. 2020. [Online]. Available: \url{https://ojs.aaai.org/index.php/AAAI/article/view/6311}
\BIBentrySTDinterwordspacing

\bibitem{journals/corr/LiMJ16a}
\BIBentryALTinterwordspacing
J.~Li, W.~Monroe, and D.~Jurafsky, ``Understanding neural networks through representation erasure.'' \emph{CoRR}, vol. abs/1612.08220, 2016. [Online]. Available: \url{http://dblp.uni-trier.de/db/journals/corr/corr1612.html#LiMJ16a}
\BIBentrySTDinterwordspacing

\bibitem{10.1145/219717.219748}
\BIBentryALTinterwordspacing
G.~A. Miller, ``Wordnet: A lexical database for english,'' \emph{Commun. ACM}, vol.~38, no.~11, p. 39–41, nov 1995. [Online]. Available: \url{https://doi.org/10.1145/219717.219748}
\BIBentrySTDinterwordspacing

\bibitem{zang-etal-2020-word}
\BIBentryALTinterwordspacing
Y.~Zang, F.~Qi, C.~Yang, Z.~Liu, M.~Zhang, Q.~Liu, and M.~Sun, ``Word-level textual adversarial attacking as combinatorial optimization,'' in \emph{Proceedings of the 58th Annual Meeting of the Association for Computational Linguistics}.\hskip 1em plus 0.5em minus 0.4em\relax Online: Association for Computational Linguistics, Jul. 2020, pp. 6066--6080. [Online]. Available: \url{https://aclanthology.org/2020.acl-main.540}
\BIBentrySTDinterwordspacing

\bibitem{jiang2023mistral}
A.~Q. Jiang, A.~Sablayrolles, A.~Mensch, C.~Bamford, D.~S. Chaplot, D.~d.~l. Casas, F.~Bressand, G.~Lengyel, G.~Lample, L.~Saulnier \emph{et~al.}, ``Mistral 7b,'' \emph{arXiv preprint arXiv:2310.06825}, 2023.

\bibitem{touvron2023llama}
H.~Touvron, L.~Martin, K.~Stone, P.~Albert, A.~Almahairi, Y.~Babaei, N.~Bashlykov, S.~Batra, P.~Bhargava, S.~Bhosale \emph{et~al.}, ``Llama 2: Open foundation and fine-tuned chat models,'' \emph{arXiv preprint arXiv:2307.09288}, 2023.

\bibitem{cai2024internlm2}
Z.~Cai, M.~Cao, H.~Chen, K.~Chen, K.~Chen, X.~Chen, X.~Chen, Z.~Chen, Z.~Chen, P.~Chu \emph{et~al.}, ``Internlm2 technical report,'' \emph{CoRR}, 2024.

\bibitem{young2024yi}
A.~Young, B.~Chen, C.~Li, C.~Huang, G.~Zhang, G.~Zhang, H.~Li, J.~Zhu, J.~Chen, J.~Chang \emph{et~al.}, ``Yi: Open foundation models by 01. ai,'' \emph{arXiv preprint arXiv:2403.04652}, 2024.

\bibitem{naik2018stress}
A.~Naik, A.~Ravichander, N.~Sadeh, C.~Rose, and G.~Neubig, ``Stress test evaluation for natural language inference,'' in \emph{Proceedings of the 27th International Conference on Computational Linguistics}, 2018, pp. 2340--2353.

\bibitem{liu2024testing}
Z.~Liu, C.~Chen, J.~Wang, M.~Chen, B.~Wu, Z.~Tian, Y.~Huang, J.~Hu, and Q.~Wang, ``Testing the limits: Unusual text inputs generation for mobile app crash detection with large language model,'' in \emph{Proceedings of the IEEE/ACM 46th International Conference on Software Engineering}, 2024, pp. 1--12.

\bibitem{li2019textbugger}
J.~Li, S.~Ji, T.~Du, B.~Li, and T.~Wang, ``Textbugger: Generating adversarial text against real-world applications,'' in \emph{Proceedings 2019 Network and Distributed System Security Symposium}.\hskip 1em plus 0.5em minus 0.4em\relax Internet Society, 2019.

\bibitem{xu2024hallucination}
Z.~Xu, S.~Jain, and M.~Kankanhalli, ``Hallucination is inevitable: An innate limitation of large language models,'' \emph{arXiv preprint arXiv:2401.11817}, 2024.

\bibitem{andriopoulos2023augmenting}
K.~Andriopoulos and J.~Pouwelse, ``Augmenting llms with knowledge: A survey on hallucination prevention,'' \emph{arXiv preprint arXiv:2309.16459}, 2023.

\bibitem{zhang2023coophance}
Q.~Zhang, Y.~Tian, Y.~Ding, S.~Li, C.~Sun, Y.~Jiang, and J.~Sun, ``Coophance: Cooperative enhancement for robustness of deep learning systems,'' in \emph{Proceedings of the 32nd ACM SIGSOFT International Symposium on Software Testing and Analysis}, 2023, pp. 753--765.

\bibitem{10041782}
Z.~Aghababaeyan, M.~Abdellatif, L.~Briand, R.~S, and M.~Bagherzadeh, ``Black-box testing of deep neural networks through test case diversity,'' \emph{IEEE Transactions on Software Engineering}, vol.~49, no.~5, pp. 3182--3204, 2023.

\bibitem{liu2021dialtest}
Z.~Liu, Y.~Feng, and Z.~Chen, ``Dialtest: automated testing for recurrent-neural-network-driven dialogue systems,'' in \emph{Proceedings of the 30th ACM SIGSOFT International Symposium on Software Testing and Analysis}, 2021, pp. 115--126.

\bibitem{10.1145/3672451}
\BIBentryALTinterwordspacing
C.~Wan, S.~Liu, S.~Xie, Y.~Liu, H.~Hoffmann, M.~Maire, and S.~Lu, ``Keeper: Automated testing and fixing of machine learning software,'' \emph{ACM Trans. Softw. Eng. Methodol.}, jun 2024, just Accepted. [Online]. Available: \url{https://doi.org/10.1145/3672451}
\BIBentrySTDinterwordspacing

\bibitem{yu2023automated}
B.~Yu, Y.~Hu, Q.~Mang, W.~Hu, and P.~He, ``Automated testing and improvement of named entity recognition systems,'' in \emph{Proceedings of the 31st ACM Joint European Software Engineering Conference and Symposium on the Foundations of Software Engineering}, 2023, pp. 883--894.

\bibitem{lyu2024keeping}
K.~Lyu, H.~Zhao, X.~Gu, D.~Yu, A.~Goyal, and S.~Arora, ``Keeping llms aligned after fine-tuning: The crucial role of prompt templates,'' in \emph{ICLR 2024 Workshop on Reliable and Responsible Foundation Models}, 2024.

\bibitem{10820047}
Y.~Huang, J.~Song, Z.~Wang, S.~Zhao, H.~Chen, F.~Juefei-Xu, and L.~Ma, ``Look before you leap: An exploratory study of uncertainty analysis for large language models,'' \emph{IEEE Transactions on Software Engineering}, vol.~51, no.~2, pp. 413--429, 2025.

\bibitem{10.1145/3691620.3695018}
\BIBentryALTinterwordspacing
Y.~Liu, J.~Yu, H.~Sun, L.~Shi, G.~Deng, Y.~Chen, and Y.~Liu, ``Efficient detection of toxic prompts in large language models,'' in \emph{Proceedings of the 39th IEEE/ACM International Conference on Automated Software Engineering}, ser. ASE '24.\hskip 1em plus 0.5em minus 0.4em\relax New York, NY, USA: Association for Computing Machinery, 2024, p. 455–467. [Online]. Available: \url{https://doi.org/10.1145/3691620.3695018}
\BIBentrySTDinterwordspacing

\bibitem{10.1145/3691620.3695001}
\BIBentryALTinterwordspacing
Q.~Zhang, C.~Zhou, G.~Go, B.~Zeng, H.~Shi, Z.~Xu, and Y.~Jiang, ``Imperceptible content poisoning in llm-powered applications,'' in \emph{Proceedings of the 39th IEEE/ACM International Conference on Automated Software Engineering}, ser. ASE '24.\hskip 1em plus 0.5em minus 0.4em\relax New York, NY, USA: Association for Computing Machinery, 2024, p. 242–254. [Online]. Available: \url{https://doi.org/10.1145/3691620.3695001}
\BIBentrySTDinterwordspacing

\bibitem{yuan2024revisiting}
L.~Yuan, Y.~Chen, G.~Cui, H.~Gao, F.~Zou, X.~Cheng, H.~Ji, Z.~Liu, and M.~Sun, ``Revisiting out-of-distribution robustness in nlp: Benchmarks, analysis, and llms evaluations,'' \emph{Advances in Neural Information Processing Systems}, vol.~36, 2024.

\bibitem{dong2023revisit}
G.~Dong, J.~Zhao, T.~Hui, D.~Guo, W.~Wang, B.~Feng, Y.~Qiu, Z.~Gongque, K.~He, Z.~Wang \emph{et~al.}, ``Revisit input perturbation problems for llms: A unified robustness evaluation framework for noisy slot filling task,'' in \emph{CCF International Conference on Natural Language Processing and Chinese Computing}.\hskip 1em plus 0.5em minus 0.4em\relax Springer, 2023, pp. 682--694.

\bibitem{liu2023robustness}
Y.~Liu, T.~Cong, Z.~Zhao, M.~Backes, Y.~Shen, and Y.~Zhang, ``Robustness over time: Understanding adversarial examples' effectiveness on longitudinal versions of large language models,'' \emph{arXiv preprint arXiv:2308.07847}, 2023.

\end{thebibliography}













\begin{IEEEbiography}[{\includegraphics[width=1in,height=1.25in,clip,keepaspectratio]{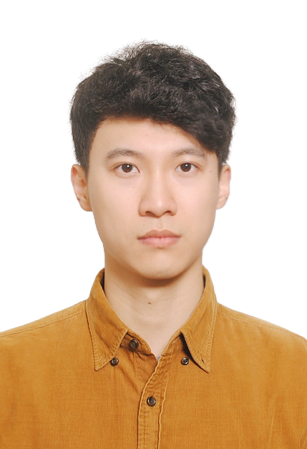}}]{Mingxuan Xiao}
is a computer science and technology student at the College of Computer Science and Software Engineering, Hohai University, currently pursuing a PH.D. degree. His current research interests include natural language processing, software engineering and metaheuristic algorithms. Before joining HHU, he received his BSc in Electronic and Information Engineering from Yancheng Teachers University.
\end{IEEEbiography}

\begin{IEEEbiography}[{\includegraphics[width=1in,height=1.25in,clip,keepaspectratio]{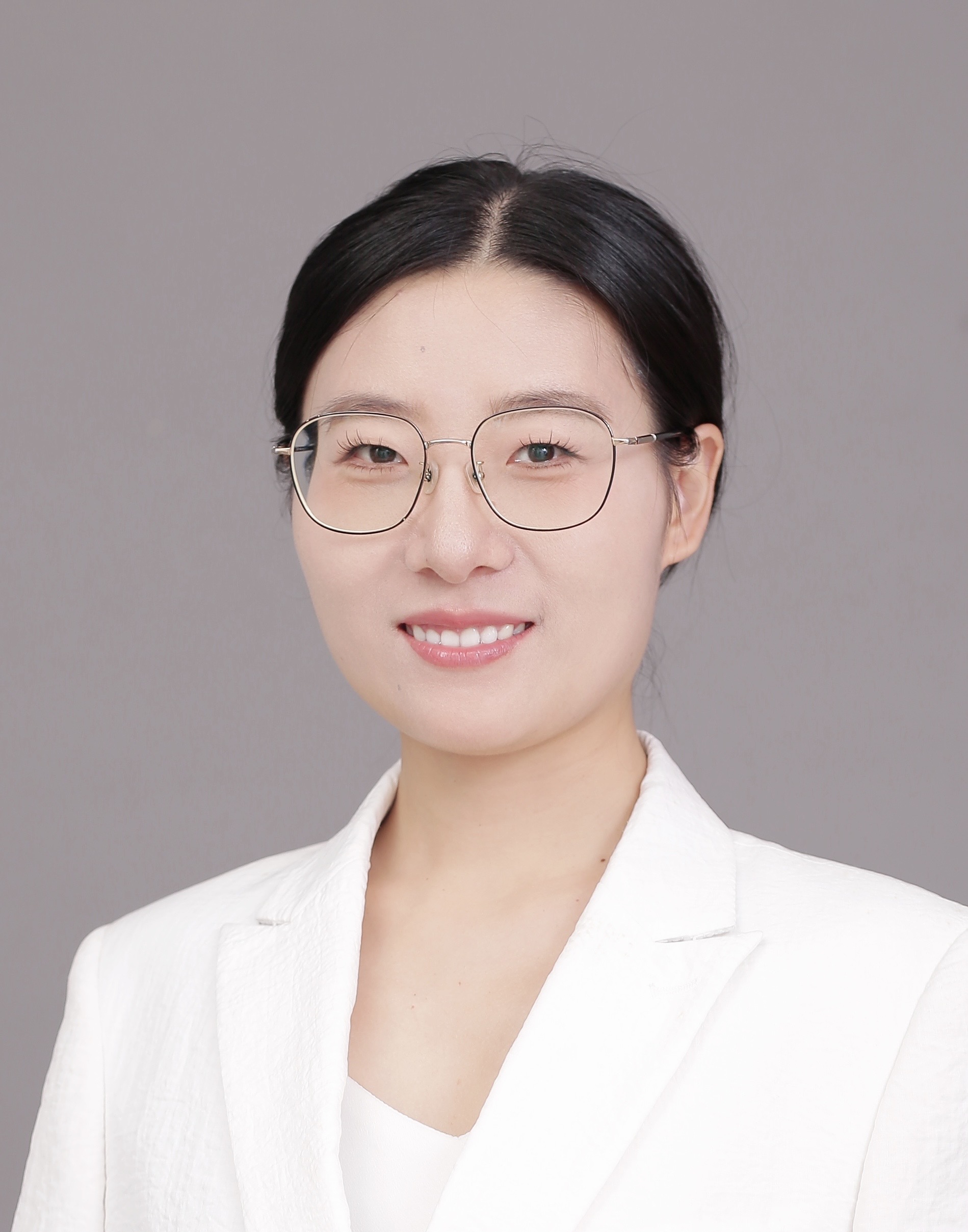}}]{Yan Xiao} is an Associate Professor at School of Cyber Science and Technology in Shenzhen Campus of Sun Yat-sen University. She received her Ph.D. degree from the City University of Hong Kong and held a research fellow position at the National University of Singapore. Her research focuses on the trustworthiness of deep learning systems and AI applications in software engineering. More information is available on her homepage: https://yanxiao6.github.io/.
\end{IEEEbiography}

\begin{IEEEbiography}[{\includegraphics[width=1in,height=1.25in,clip,keepaspectratio]{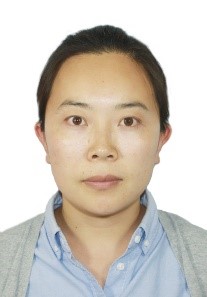}}]{Shunhui Ji} received the B.S. degree in computer science and technology and the Ph.D. degree in computer software and theory from Southeast University,
in 2008 and 2015, respectively. She is currently an
Associate Professor with the College of Computer
and Information, Hohai University, Nanjing, China.
Her research interests include service computing,
cloud computing, software modeling, analysis, testing,
and verification. She is a Reviewer of some
international conferences and journals.
\end{IEEEbiography}

\begin{IEEEbiography}[{\includegraphics[width=1in,height=1.25in,clip,keepaspectratio]{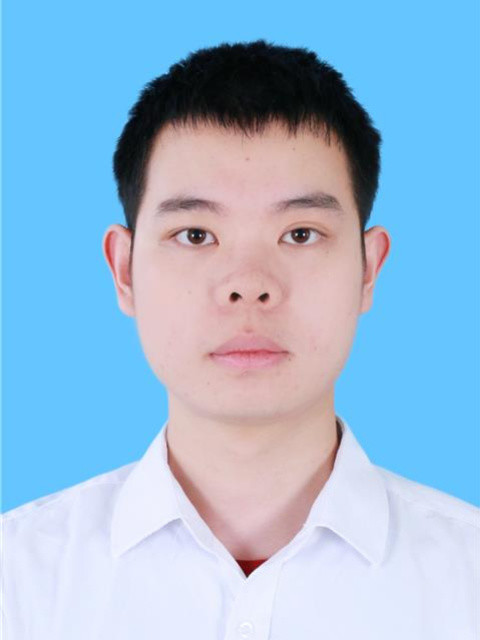}}]{Hanbo Cai} received the M.Eng. degree in software engineering from Guangxi Normal University in 2021. He is currently pursuing the Ph.D. degree in computer science and technology with the College of Computer Science and Software Engineering, Hohai University. His research interests include the domain of trustworthy ML and responsible AI, especially backdoor learning and adversarial attacks.
\end{IEEEbiography}

\begin{IEEEbiography}[{\includegraphics[width=1in,height=1.25in,clip,keepaspectratio]{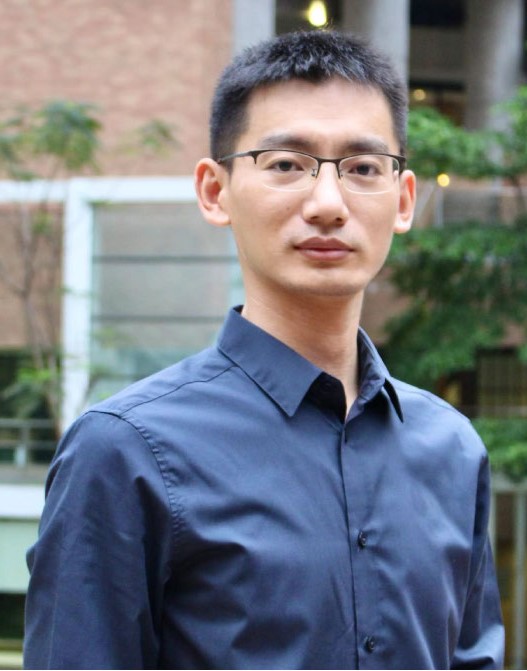}}]{Lei Xue} received the Ph.D. degree in computer science from The Hong Kong Polytechnic University.
He is an Associated Professor with the School of
Cyber Science and Technology, Sun Yat-sen University. His current research topics mainly focus on
mobile and IoT system security, program analysis,
and automotive security.
\end{IEEEbiography}

\begin{IEEEbiography}[{\includegraphics[width=1in,height=1.25in,clip,keepaspectratio]{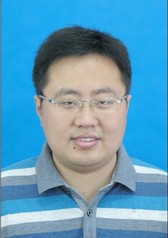}}]{Pengcheng Zhang}
received the Ph.D. degree in computer science from Southeast University in 2010. He is currently a full professor in College of Computer Science and Software Engineering, Hohai University, Nanjing, China, and
was a visiting scholar at San Jose State University, USA. His research interests include software engineering, service computing and data science. He has published research papers in premiere or famous computer science journals, such as IEEE Transactions on Software Engineering, IEEE Transactions on Services Computing,
IEEE Transactions on Knowledge and Data Engineering, IEEE Transactions on Big Data, IEEE Transactions on Emerging Topics in Computing, IEEE Transactions on Cloud Computing  and IEEE Transactions on Reliability. He was the co-chair of
IEEE AI Testing 2019 conference. He served as a technical program committee member on various international conferences. He is a member of the IEEE.
\end{IEEEbiography}

\vfill

\end{document}